\documentclass[10pt,twocolumn,preprintnumbers,amsmath,amssymb,nofootinbib,superscriptaddress]{revtex4-1}

\usepackage{epsfig}
\usepackage{url}
\usepackage[normalem]{ulem}
\usepackage{latexsym}
\usepackage{epsfig}
\usepackage{amsmath}
\usepackage{amssymb}
\usepackage{wasysym}
\usepackage{graphicx}
\usepackage{verbatim}
\usepackage{enumerate,mdwlist}
\usepackage[titletoc]{appendix}
\usepackage{amsfonts}
\usepackage{pgfplots}
\usepackage[export]{adjustbox}
\usepackage{bbold}
\usepackage[normalem]{ulem}

\usepackage[linktocpage=true]{hyperref}
\hypersetup{
colorlinks=true,
citecolor=ultramarine,
linkcolor=cadmiumgreen,
urlcolor=indigo(dye),
pdfauthor={},
pdftitle={},
pdfsubject={}
}

\usepackage[all]{xy} 
\usepackage{amsfonts}

\newcommand{\ba}{\begin{eqnarray}}
\newcommand{\ea}{\end{eqnarray}}
\newcommand{\be}{\begin{equation}}
\newcommand{\ee}{\end{equation}}

\newcommand{\gwphaseurl}{https://github.com/ezquiaga/phazap}
\newcommand{\gwphaselink}{\href{\gwphaseurl}{\texttt{\gwphaseurl}}}
\newcommand{\gwphase}{\href{\gwphaseurl}{\texttt{phazap}}}

\newcommand{\phiH}{\phi_\mathrm{H}}
\newcommand{\phiL}{\phi_\mathrm{L}}
\newcommand{\phiV}{\phi_\mathrm{V}}
\newcommand{\tauHL}{\tau_\mathrm{HL}}
\newcommand{\tauHV}{\tau_\mathrm{HV}}
\newcommand{\tauLV}{\tau_\mathrm{LV}}
\newcommand{\HV}{\mathrm{HV}}
\newcommand{\HL}{\mathrm{HL}}

\newcommand{\BLU}{B_\mathrm{U}^{\mathrm{L}}}
\newcommand{\CLU}{C_\mathrm{U}^{\mathrm{L}}}

\newcommand{\fref}{f_\mathrm{ref}}
\newcommand{\rref}{\mathrm{ref}}
\newcommand{\ra}{\mathrm{ra}}
\newcommand{\dec}{\mathrm{dec}}

\definecolor{grey}{rgb}{0.4,0.4,0.4}
\definecolor{dullmagenta}{rgb}{0.4,0,0.4}
\definecolor{darkblue}{rgb}{0,0,0.4}
\definecolor{midblue}{rgb}{0,0,0.5}
\definecolor{midred}{rgb}{0.5,0,0}
\definecolor{orange}{rgb}{1,0.5,0}
\definecolor{lightbrown}{rgb}{0.75,0.5,0.25}
\definecolor{tan}{cmyk}{0.14,0.42,0.56,0}
\definecolor{djunglegreen}{cmyk}{0.99,0,0.52,0}
\definecolor{lightgreen}{rgb}{0,1,0}
\definecolor{olivegreen}{cmyk}{0.64,0,0.95,0.40}
\definecolor{midgreen}{rgb}{0.0,0.675,0.0}
\definecolor{darkgreen}{rgb}{0,0.5,0}
\definecolor{ultramarine}{rgb}{0.07, 0.04, 0.56}
\definecolor{cadmiumgreen}{rgb}{0.0, 0.42, 0.24}
\definecolor{indigo(dye)}{rgb}{0.0, 0.25, 0.42}

\begin{document}

\title{Identifying strongly lensed gravitational waves through their phase consistency}

\author{Jose Mar\'ia Ezquiaga}
\email{jose.ezquiaga@nbi.ku.dk}
\affiliation{Niels Bohr International Academy, Niels Bohr Institute, Blegdamsvej 17, DK-2100 Copenhagen, Denmark}

\author{Wayne Hu}
\email{whu@background.uchicago.edu}
\affiliation{Kavli Institute for Cosmological Physics, Department of Astronomy \& Astrophysics, Enrico Fermi Institute, The University of Chicago, Chicago, IL 60637, USA}

\author{Rico K. L. Lo}
\email{kllo@caltech.edu}
\affiliation{LIGO, California Institute of Technology, Pasadena, CA 91125, USA}

\begin{abstract}
Strongly lensed gravitational waves (GWs) from binary coalescence manifest as repeated chirps from the original merger. At the detectors, the phase of the lensed GWs and its arrival time differences will be consistent modulo a fixed constant phase shift. We develop a fast and reliable method to efficiently reject event pairs that are not-lensed copies and appropriately rank the most interesting candidates. Our method exploits that detector phases are the best measured GW parameter, with errors only of a fraction of a radian and differences across the frequency band that are better measured than the chirp mass. The arrival time phase differences also avoid the shortcomings of looking for overlaps in highly non-Gaussian sky maps. Our basic statistic determining the consistency with lensing is the distance between the phase posteriors of two events and it directly provides information about the lens-source geometry which helps inform electromagnetic followups. We demonstrate that for simulated signals of not-lensed binaries specifically chosen with many coincident properties so as to trigger false lensing alarms none of the pairs have phases closer than $3\sigma$, and most cases reject the lensing hypothesis by $5\sigma$. Looking at the latest catalog, GWTC3, we find that only $6\%$ of the pairs are consistent with lensing at 99\% confidence level. Moreover, we reject about half of the pairs that would otherwise favor lensing by their parameter overlaps and demonstrate good correlation with detailed joint parameter estimation results. This reduction of the false alarm rate will be of paramount importance in the upcoming observing runs and the eventual discovery of lensed GWs. Our code is publicly available and could be applied beyond lensing to test possible deviations in the phase evolution from modified theories of gravity and constrain GW birefringence.
\end{abstract}

\date{\today}

\maketitle

\section{Introduction}
Ground-based gravitational-wave (GW) observatories coherently detect the space-time perturbations produced by merging compact objects such as binary black holes. The phase evolution of these signals encode information about their gravitational properties, astrophysical origin and cosmological propagation. The LIGO--Virgo--KAGRA (LVK) \cite{Aasi2015,Acernese_2014,KAGRA:2020agh} detectors have already accumulated observations of about a hundred compact binaries during the first three observing runs \cite{LIGOScientific:2021djp}. 

Compact binary coalescence signals are inevitably affected by the intervening matter along their travel path. 
The effect of such gravitational interaction is typically negligible. However, for sufficient alignment between the source and the matter distribution, the latter acts as a lens magnifying and possibly distorting the original signal. 
For large lenses such as galaxies or clusters of galaxies, strong lensing effects can lead to nearly identical chirps of the same signal arriving at the detectors days to months apart. 
The probability of observing strongly lensed GWs depends heavily on the source and lens populations, but for galaxy lenses rates can be up to 1 lensed event in every 1000 events \cite{Ng:2017yiu,Xu:2021bfn,Wierda:2021upe}. 
As we expect hundreds to thousands of mergers within the fourth and fifth observing runs, the first GW lensing detection could be within the coming years. 

Confidently identifying strongly lensed GWs is however a challenging task. Each of the repeated instances of the lensed signals hide within large catalogs in which not-lensed events can easily mimic lensed ones \cite{Caliskan:2022wbh}. 
Past strongly lensed searches \cite{Hannuksela:2019kle,LIGOScientific:2021izm,LIGOScientific:2023bwz,Janquart:2023mvf} have focused on computing the overlap between parameters of different events in order to account for the likelihood of lensing \cite{Haris:2018vmn} before launching a more computationally costly joint parameter estimation \cite{Janquart:2021qov,Lo:2021nae}. 
However, with current sensitivities, the physical parameters describing the binary are not well constrained and degenerate among each other, making it common to find large overlaps in not-lensed pairs.    
Therefore, in order not to miss any lensed event, many false alarms would have to be followed up. 

In this work we develop a fast and reliable method to identify strongly lensed candidates through the consistency of the best measured GW quantities, the phases at the detectors. 
We borrow some of the tension statistics developed in cosmology \cite{Raveri:2018wln} to compute the confidence level (CL) of agreement between two events. 
The advantage of our method is that it establishes a well defined measure of how inconsistent (in ``tension'') a pair of GWs is with the strong lensing hypothesis. Thus, by construction, our method 
generates a highly complete catalog of lensing candidates while also reducing the number of false alarms compared to previous methods. 
A code that implements our method, \gwphase, is publicly available.

We begin the paper in Sec.\ \ref{sec:GWobservables} showing how to efficiently reconstruct the main GW observables at the detectors --- detector phases, time delay phases and polarization states, from pre-existing parameter estimation posteriors. 
In Sec.\ \ref{sec:strong_lensing} we then show how those observables are affected by strong lensing. 
In Sec.\ \ref{sec:Method} we outline our method to analyze lensed candidates. 
We apply the method to a set of simulated events and to the latest LVK catalog in Secs.\ \ref{sec:simulatedevents} and \ref{sec:gwtc} respectively. 
Our results are compared with other methods in Sec.\ \ref{sec:comparison}. 
We then explore the implications for current and future strong lensing searches in Sec.\ \ref{sec:strong_lensing_searches}.  
We conclude in Sec.\ \ref{sec:conclusions} with the main results and possible extensions of our method including those beyond lensing.
In a series of appendices we provide further details on our method and conventions. In particular, we detail our reference frame conventions, demonstrate how to compute the phase at a new frequency, show how to break the sky localization bimodalities, estimate the errors of the detector phases, specify the settings of the simulated GW events and report the detector phases for the real GWs analyzed in Apps.\ \ref{sec:frames}, \ref{sec:phase_new_freq}, \ref{app:breaking_bimodality}, \ref{sec:leading_order}, \ref{app:injections} and \ref{app:gwtc} respectively.

\section{Gravitational wave observables}
\label{sec:GWobservables}
The gravitational wave emission of a compact binary coalescence depends on the intrinsic parameters of the source, such as the masses ($m_i$) and spin vectors ($\vec{S}_i$), and can be described by the two tensorial polarizations $h_+$ and $h_\times$. 
The detected GW strain $h_d(t)$ at each detector also depends on the extrinsic parameters: the distance to the source $d_L$, its sky location as determined by the line of sight $\vec{n}=\{\ra,\dec\}$, the polarization orientation $\psi$ and arrival time $t_d$, the  orbital inclination $\iota$ and phase $\phi_\rref$ at a reference frequency $\fref$:
\begin{align}
    h_d(t-t_d) &= F_+h_+ + F_\times h_\times\nonumber\\
    &= \int^\infty_{-\infty}\tilde{h}_d(f)e^{2\pi if(t-t_d)}df,
        \label{eq:detectedwaveform}
\end{align}
where $F_{+,\times}=F_{+,\times}(\vec{n},\psi,t_d)$ are the antenna pattern functions, $\tilde{h}_d(f,m_{1,2},\vec{S}_{1,2},\iota,d_L,\phi_\rref)$ is the complex-valued Fourier transform of the real-valued function $h_d(t)$, and the subscript $d$ labels quantities evaluated at each detector (see App.\ \ref{sec:frames} for conventions on the different reference frames). 
For the rest of the paper we will focus on the positive frequency modes, whose negative frequency counterparts can be computed by the reality condition $\tilde h(f)=\tilde h^*(-f)$.

Before projecting into the detector frame, we can further decompose the frequency domain signal of a given polarization in the radiation frame into a sum of frequency-dependent amplitudes and phases for each multipole mode 
at emission. For cases with precession, this is usually done in a coprecessing frame and there the dominant mode is $l=|m|=2$ which we will refer to as the 22-mode. 
Higher order modes are only sizable when the binary has highly asymmetric component masses or eccentricity. 
On the other hand since precession and higher modes are part of the standard analysis they impact parameter degeneracies and inference regardless of whether they are detected in a given event or not.
For our analysis, we will therefore use  waveform models containing these effects. 

From the observation of the amplitudes, phases and arrival times at each detector one can reconstruct the properties of the detected signal with Bayesian parameter estimation. Despite the precise measurement of the frequency-dependent phase  of the signal, the high-dimensionality of the parameter space of compact binary coalescences, 15 dimensions for quasi-circular binaries accounting for 8 intrinsic $\{m_1,m_2,\vec{S}_1,\vec{S}_2\}$ and 7 extrinsic $\{\ra,\dec,\psi,\iota,d_L,t_\rref,\phi_\rref\}$ parameters, leads to poorly constrained marginalized posteriors for each of the parameters individually and complicated joint posteriors due to their degeneracies, see e.g.\ \cite{Roulet:2022kot}. 
Still, one can computationally efficiently use the \emph{full} parameter estimation to \emph{derive} the well constrained posterior probability for the phases of the 22-mode at every frequency and detector, which we hereafter refer to as the ``detector phases": 
\begin{equation}\label{eq:phi_d}
    \phi_{d}(f)=\phi_{22}(f)+\chi^d_{22}(\vec{n},\psi',\iota',t_d)\,,
\end{equation}
where $\phi_{22}(f)$ is the global phase of the 22-mode,\footnote{Specifically, for precessing binaries this is associated to the phase of the 22-mode in the co-precessing $\vec L$-frame. With our conventions and exploiting the equatorial symmetry, this is built from the phase of the co-precessing $l=2$, $m=-2$, $+$-polarization mode for positive frequencies.}
$\chi_{lm}=\text{arctan}[F_+,a_{lm}(\iota')F_\times]$, 
and 
$a_{lm}=(1-r_{lm})/(1+r_{lm})$ with the ratio of the circular polarization amplitudes $r_{lm}=|h_L^{lm}|/|h_R^{lm}|$, see App.\ \ref{sec:phase_new_freq} for technical details on how to derive this expression with a waveform-based decomposition of the polarization state. 
For the 22-mode: $a_{22}(\iota)=2\cos \iota/(1+\cos^2 \iota)$. 
This defines the phase of the  22-component of the detector strain at frequency $f$  at the arrival time 
$t=t_d$. 
Note that precession causes a frequency-dependent rotation of the linear polarization angle, $\psi'=\psi+\zeta(f)$, as well as a frequency-dependent inclination angle $\iota'=\iota(f)$ or circular polarization state. 

\begin{figure}[t!]
\centering
\includegraphics[width = \columnwidth,valign=t]{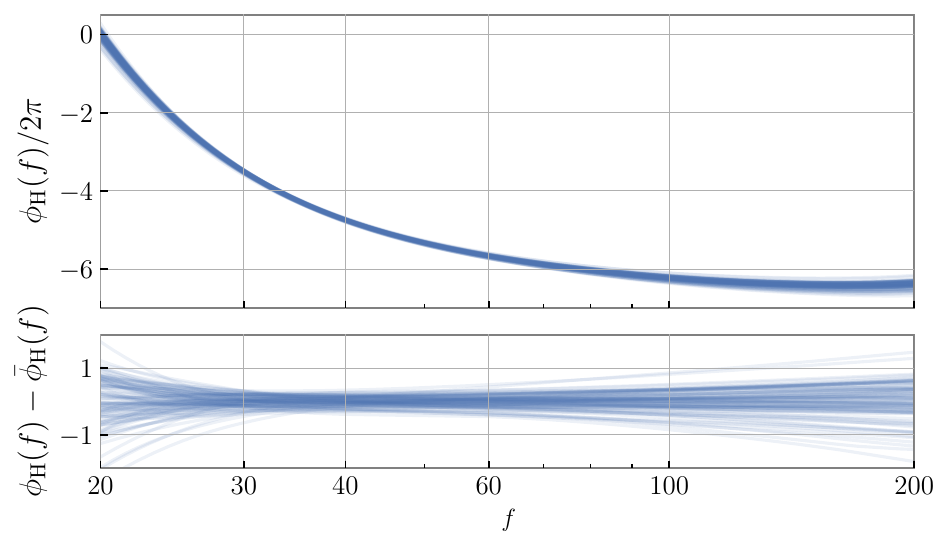}
\caption{
Reconstructed detector phase as a function of frequency for the first detected event, GW150914 \cite{LIGOScientific:2016aoc}, in the Hanford (H) detector. In this frequency range, the phase runs approximately over six cycles with errors of a fraction of a radian with respect to the mean evolution $\bar\phi_\mathrm{H}(f)$. Data is from the latest GWTC-2.1 catalog \cite{LIGOScientific:2021usb}.}
\label{fig:phi_H_fs}
\end{figure}

The error in the detected phase scales inversely with the signal-to-noise (SNR) at the detector, $\delta\phi_d \sim 1 / \rho_d$ \cite{Finn:1992xs,Cutler:1994ys}. (see App.\ \ref{sec:leading_order} for a simple derivation).  
While the phase at a given frequency will give us an absolute phase that will be important for the lensing consistency, the phase difference $\Delta\phi_{f,d}=\phi_d(f)-\phi_d(f')$ will be relevant to assess the orbital phase evolution, 
accounting for the number of cycles, which encodes information about the masses of the binary through the detector frame chirp mass $\mathcal{M}_c=(m_1m_2)^{3/5}/(m_1+m_2)^{1/5}$. 

In Fig.\ \ref{fig:phi_H_fs} we show the reconstructed phase as a function of frequency at a given detector for GW150914, the first GW \cite{LIGOScientific:2016aoc} using data from the latest analysis \cite{LIGOScientific:2021usb}. The errors in the phase are a fraction of a radian over a large range of frequencies. 
Importantly, as we will show later, the phase difference is a better discriminator than the chirp mass due to its smaller relative error and larger range that reduces the probability of consistency by random chance. 
In addition, it can be seen that the best measured phase is around 40Hz, higher than the fiducial reference frequency of 20Hz but lower than the frequency where the amplitude peaks, $\sim 150$Hz \cite{LIGOScientific:2016aoc}. This is because at lower frequencies there are more cycles and larger SNR is accumulated as long as the signal is in band. 
At 40Hz the standard deviation in the phase at the Hanford detector $\phiH$ is $0.15$ radians, or 
 $\delta\phiH({\rm 40 Hz}) \sim 3 / \rho_\mathrm{H}$. Note that this order unity factor multiplying the $1/\rho_d$ scaling can vary from event to event but we will use $\sim 3$ as a rough guide for estimation purposes later.
In large part, this variation is due to events with high masses and short in-band durations where the 22-phase becomes hard to distinguish especially in the presence of other parameters (see App.~\ref{sec:leading_order}). 

\begin{figure}[t!]
\centering
\includegraphics[width = \columnwidth,valign=t]{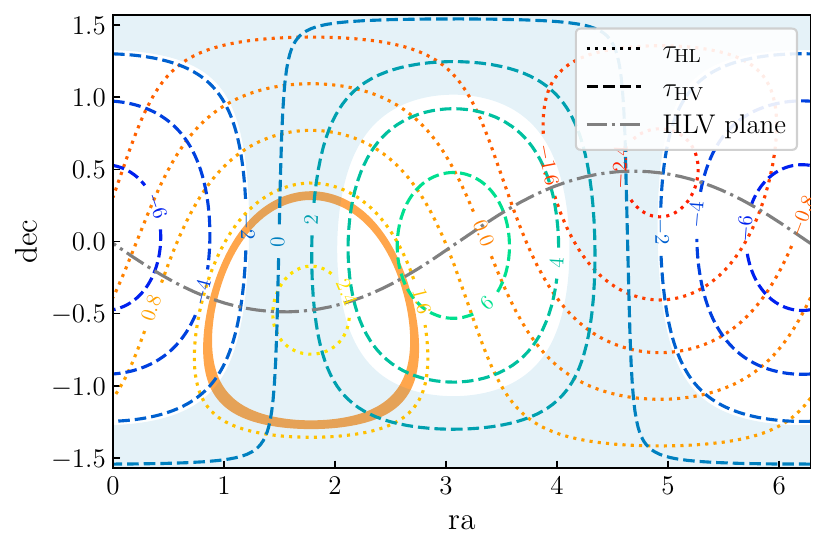}
\caption{
Time delay phase contour lines as a function of the right ascension (ra) and declination (dec) for GW150914 \cite{LIGOScientific:2016aoc}. 
Dotted lines indicate contours for Hanford and Livingston ($\tauHL$), while dashed lines are for Hanford and Virgo ($\tauHV$). 
The shaded orange and blue regions correspond to the $95\%$ CL from the reconstructed time delays, $\tauHL$ and $\tauHV$ respectively. 
The dash-dotted line indicates the plane defined by the position of the HLV detectors.  
The intersection of both shaded regions occurs in reflection symmetric positions above and below the detector’s plane and correspond to a bimodality in the localization from time delays.
}
\label{fig:time_delay_contours}
\end{figure}

In addition to the phase, detected GWs have timing information, which is used to triangulate the sky position of the source. 
Typically, in the parameter estimation, the arrival time of the signal $t_c$ is defined by the arrival of the maximum of the time domain strain, $\mathcal{A}(t)^2=|h_+|^2+|h_\times|^2$, as measured at a common GPS time in the Earth frame. 
The maximum of the signal is chosen to approximately track the coalescence time of the binary. 
We can reconstruct the arrival time at each detector as
\begin{equation}
    t_d = t_c - \vec{n}\cdot\vec{r}_d/c\,,
\end{equation}
where $\vec{r}_d$ is the position of each detector. 
We use the relative arrival time at different detectors to define the additional phases 
\begin{equation}\label{eq:tau_d1d2}
    \tau_{d_1d_2}(f)\equiv2\pi f{\Delta t_{d_1d_2}}= 2\pi f 
    \vec{n}\cdot \vec{r}_{d_2 d_1} /c\,,
\end{equation}
where $\vec{r}_{d_2 d_1 } \equiv \vec{r}_{d_2}-\vec{r}_{d_1}$. 
The advantage of these time delay phases is that they largely remove degeneracies in localization as we shall now discuss. 
In general, the arrival time difference between two detectors constrains one angle, $\vec{n}\cdot \vec{r}_{d_1 d_2}$, defining a ring in the sky of possible source locations. 
With three detectors, two angles are constrained, $\vec{n}\cdot \vec{r}_{d_2 d_1}$ and $\vec{n}\cdot\vec{r}_{d_1 d_3}$, defining two rings in the sky that intersect at two points where the event localization is possible.  
These two possible localization regions correspond to a reflection symmetry of the time delays across the hemispheres delineated by the plane defined by the position of the three detectors, i.e.\  distinguished by the sign of $\vec{n}\cdot (\vec{r}_{d_1 d_2} \times \vec{r}_{d_1 d_3})$. 
As an example, we plot the time delay phase contour lines of $\tauHL$ and $\tauHV$ as a function of right ascension and declination for GW150914 in Fig.\ \ref{fig:time_delay_contours}. The shaded regions corresponds to the $95\%$ CL for the reconstructed time delay phases and their intersection indicates the possible sky localization above and below the plane of the detectors.
Note that GW150914 was not measured by Virgo and thus $\tauHV$ is derived from the localization supplied by the two LIGO detectors rather than measured directly.  
For sufficiently high SNR, the amplitude and polarization information enters into the localization and this information breaks the time delay reflection symmetry.  
In detector networks with more than three detectors, this degeneracy is also broken and 
more than two time delay phases can be computed. On the other hand, these multiple phases always redundantly  parameterize the two sky angles. 

\begin{figure*}[t!]
\centering
\includegraphics[width = 0.49\textwidth,valign=t]{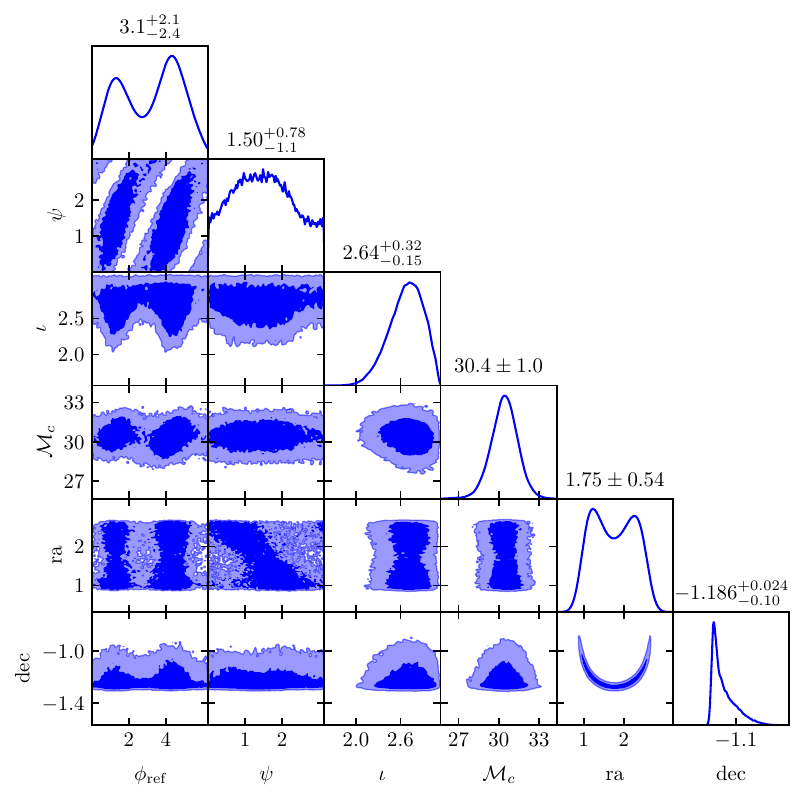}
\includegraphics[width = 0.49\textwidth,valign=t]{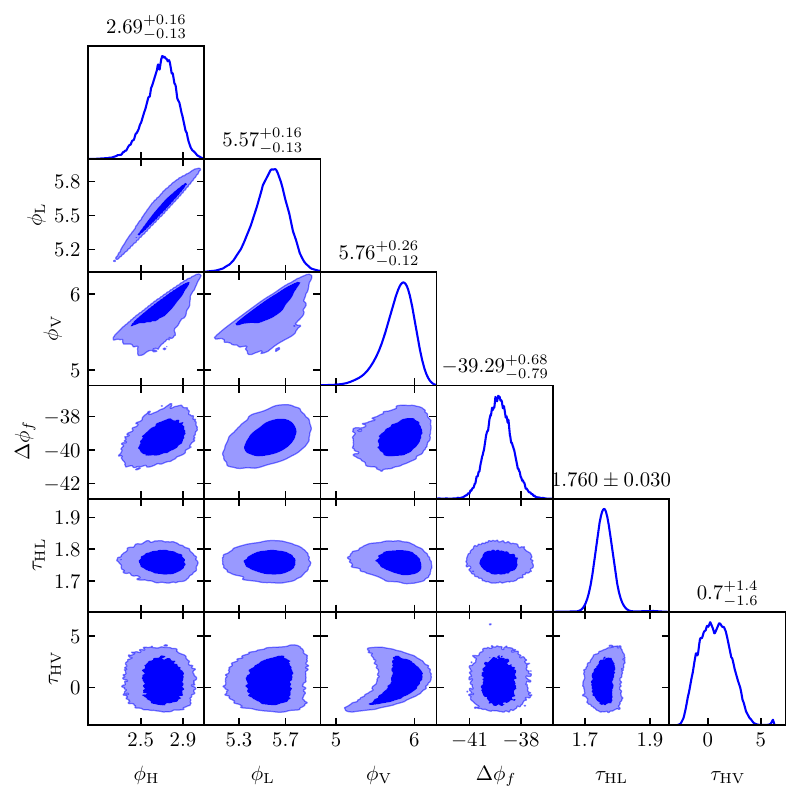}
\caption{
Comparison between the waveform parameters and the reconstructed phases for GW150914. On the left, the posterior distributions reference phase at the 20 Hz reference frequency  ($\phi_\text{ref}$), the polarization angle ($\psi$), the inclination at the reference frequency ($\iota$), the right ascension (ra), the declination (dec) and the detector frame chirp mass ($\mathcal{M}_c$). 
On the right, the detector phases ($\phi_\mathrm{H,L,V}$) and the time delay phases ($\tau_\mathrm{HL,HV}$) at our pivot frequency of 40Hz, and the phase evolution between 20 and 100Hz ($\Delta\phi_f$). 
In both plots the contour lines indicate the 68 and $95\%$ credible intervals, while the quoted ranges in the titles are the median and $68\%$ credible interval. 
}
\label{fig:phases_vs_pe_parameters}
\end{figure*}

Finally, the amplitude information in GW events carries information beyond the localization.  
The overall amplitude serves to quantify the distance to the source and the lensing magnification while the relative amplitudes between detectors help fix the polarization state through the polarization  angles $\beta$ and $\zeta$, respectively (see App.\ \ref{sec:phase_new_freq} for details). 
As with the time delay phases, by deriving the detector phases from the parameter posteriors rather than directly measuring them in each detector, we encode some of this extra information on the polarization state into the inferred parameters.

In summary, for the current network of three detectors LIGO Hanford (H), LIGO Livingston (L) and Virgo (V), we can construct six parameters: three detector phases $\phiH,\phiL,\phiV$ and two arrival-time phases $\tauHL$and $\tau_{LV}$ by evaluating (\ref{eq:phi_d}) and (\ref{eq:tau_d1d2}) at a given frequency, as well as the orbital phase evolution parameter $\Delta\phi_f$ 
(our default choice is to reconstruct this phase difference at H, $\Delta\phi_f\equiv \Delta\phi_{f,\mathrm{H}}$, although the detector dependence of the  difference is minimal when derived from the full parameter estimation).  

This is to be compared with the five extrinsic parameters $\ra,\dec,\psi,\iota,\phi_\rref$ and the intrinsic parameter, the detector frame chirp mass $\mathcal{M}_c$ (as well as the remaining of the original 15 parameters which are marginalized over). 
By inferring constraints on the former set from the latter set, we efficiently encapsulate the information from parameter estimation into quantities that are closer to the direct observables and hence have reduced degeneracies and multimodality. 

To exemplify this transformation, a plot of the reconstructed phases compared to the original parameters for GW150914 is presented in Fig.\ \ref{fig:phases_vs_pe_parameters}. 
Here one can also see that the relative error in the orbital phase evolution $\Delta\phi_f$ is smaller than in the chirp mass. 
Moreover, the errors in the $\phiH$ and $\phiL$  are similar as the SNR in both detectors is comparable and information is shared in the reconstruction, whereas their difference $\phiH-\phiL$ is even better constrained. 
Similarly despite GW150914 having no Virgo data, the phase that Virgo would have seen, $\phiV$, is still constrained from the parameter estimation, although not as well as $H$ and $L$ due to parameter degeneracies that are not broken by direct measurement. 
Likewise the  time delay $\tauHV$ is also constrained by inference. 
For this loud event the full sky localization degeneracy is partially broken and the  $95\%$ localization region contains only the lower intersection region in Fig.\ \ref{fig:time_delay_contours} contributes.  
This also helps in having a well constrained reconstructed phase in Virgo. 
However, in lower SNR events, as we will see later, the phase in the detectors not online is typically unconstrained.  

Overall, though, the better constraints using our derived detector phase parameters vs.~the original binary parameters  illustrated in Fig.\ \ref{fig:phases_vs_pe_parameters} reflects the advantage of our approach in capturing the GW observable and carries over to lensing identification as we shall see next.

\section{Gravitational wave strong lensing}\label{sec:strong_lensing}
In the regime of strong lensing, multiple instances of the same event are produced, each of them acquiring a change in amplitude, arrival time and phase \cite{Schneider:1992}
\begin{equation}
\begin{split}
    &\tilde{h}_\mathrm{lensed}^{j}(f)=\sqrt{|\mu_j|}e^{in_j\pi/2}\tilde{h}(f)\,, \\
    &t_d^j = t_d + \Delta t_j\,
\end{split}
\end{equation}
with $n_j=0,1,2$ for type I, II and III images respectively and $\Delta t_j$ as the time delay compared to the signal without lensing.
Therefore, if the detectors were in the same positions relative to the source for each image (see \S \ref{sec:Method}), the detector phase of two lensed images $j$ and $k$ should only differ by their Morse phase
\begin{equation}
    \Delta \phi^d_{jk} = (n_j-n_k)\pi/2\,,
\end{equation}
meaning that their phase should be identical if they are the same image type or differ by a multiple of $\pi/2$ otherwise. 
Distances to the source and arrival times will be biased by the magnification and time delay and thus cannot be used to reject the lensing hypothesis in a given pair of events, though with additional astrophysical assumptions both can be informative. 

In the regime of geometric optics, the polarization rotation is of the order of the deflection angle and, as a consequence, negligible for ground-based detectors \cite{Ezquiaga:2020gdt}. All instances of the original event should then have the same polarization state as a function of frequency. 
For example, the degree of circular polarization can be parametrized by
\begin{equation}
    r=|h_L|/|h_R|\,,
\end{equation}
the ratio of the left- and right-handed polarization amplitudes (see App.\ \ref{sec:phase_new_freq} for other quantities defining the polarizations). 
As discussed in the previous section, this information is encoded in the detector phases inferred from parameter estimation. 
Future observations of the polarization states could be used, for instance, to test the spin consistency among the lensed events. 

Altogether, in order to identify strongly lensed GWs we can look at the consistency of the detector phases and time delays.
For the current network of detectors, the polarization state is not separately well constrained and for that reason we do not consider its consistency directly. 
Focusing on the GW detector phases is advantageous over individual waveform parameters since their large degeneracies and measurement uncertainties make them easily overlap by chance, leading to high false alarm rates \cite{Caliskan:2022wbh,Wierda:2021upe,Janquart:2021qov}. 
This is evident, for example, when looking at Fig.\ \ref{fig:phases_vs_pe_parameters}.
Previous analyses have ranked lensed candidates by computing overlaps in the redshifted component masses, spins and sky positions  as we will discuss in Sec.\ \ref{sec:comparison}. 

It is also important to note that for this consistency test to work, the standard parameter estimation should use models that are a good description of the lensed signals. This is the case for the type I and III images, but the phase shift of type II images could induce waveform distortions that are not included within the family of (not-lensed) general relativity waveforms \cite{Ezquiaga:2020gdt}. Therefore, for type II images the parameter estimation of sky positions and inclinations could be biased \cite{Vijaykumar:2022dlp}. 
However, we have checked with simulated type II images, where the Morse phase shift is applied directly to each frequency, the detected phase is well recovered and consistent with the expected signal-to-noise ratios of current detectors. 
Extraordinarily loud type II images, specially when having asymmetric masses in close to edge on binaries \cite{Ezquiaga:2020gdt,Wang:2021kzt,Janquart:2021nus}, could be identified directly by including the phase factor in the parameter estimation. 
We expect similar results to hold for strongly lensed images that are also lensed by smaller lenses inducing waveform distortions: if the distortion is large enough to bias the parameter estimation, it could be identified as lensed by targeted searches.

\section{Method}
\label{sec:Method}

Our goal is to design a fast and reliable method that rejects  as many non-lensed events as possible while selecting the most promising events for more detailed study. 
For that reason, instead of re-analyzing each candidate pair using joint parameter estimation, we efficiently post-process the original posterior samples from the parameter estimation that are performed for every detection. 

We analyze the events of a given GW catalog in pairs, aiming to first discard those pairs whose phases are not compatible with the strong lensing hypothesis and then rank the remaining candidates for priority in a joint parameter estimation campaign. 
We generically label the events in the pair as ``event 1" and ``event 2".  
To compare the detector phases of two events, we need first to fix the reference frequency and establish a common detector reference frame. 

We fix the reference frequency to $40$Hz, which for planar, quasi-circular inspairaling binaries and advanced LIGO and Virgo detectors approximately gives the best measured phase. Note that this choice is different from the standard $\fref=20$Hz of LVK catalogs, but it is easy to compute the detector phases at a new frequency: see App.\ \ref{sec:phase_new_freq}. Unlike  Ref.\ \cite{Roulet:2022kot}, we cannot optimize this frequency per event since we need a common parameterization between pairs.  Still, if a more optimal strategy is desired, one could alternately choose the frequency for which the joint errors are minimized at the computational expense of re-processing per pair. 
As a bonus, at the best measured frequency, the detected phases de-correlate from other intrinsic parameters such as the chirp mass. This is not in general the case at other frequencies.   We defer such optimizations to future studies.

In place of the chirp mass we take the phase difference across the widest frequency range where the detected phase is well constrained. Our results are not very sensitive to the precise value since the larger phase errors of a larger range are compensated by the increase in the number of cycles. 
Our default choice is from 20 to 100Hz, which matches the well-constrained region of  the fiducial case displayed in Fig.\ \ref{fig:phi_H_fs}.  
Similarly, we compute the orbital phase evolution at Hanford, although the detector's dependence essentially drops out.

As discussed in the previous section, the detector phases between the two lensed events will differ by multiples of $\pi/2$ only if the detectors are in a common reference frame between events. 
We fix the reference frame to the arrival of event 1. For event 2, this can be achieved by shifting the mean of the arrival of event 2:
\begin{equation}
    t_2\to t_2 - \langle t_2\rangle + \langle t_1\rangle\,. 
\end{equation}
Note that this changes the detector phases of event 2  by changing $\chi_d$ in Eq. (\ref{eq:phi_d}). 
With this choice we are comparing the actual ``detected phases" of event 1 with the inferred ``detector phases" of event 2, defined by what the detectors would have seen, had they been in the orientations of event 1 given the parameter estimation of event 2.  
The main caveat in this approach is that the detector phases of event 2 will generally have poorer constraints, larger degeneracies, and less Gaussian distributed posteriors than the detected phases of event 2. 

Likewise, the time delay phases for event 2 reflect only the localization of event 2 through Eq.~(\ref{eq:tau_d1d2}) and do not directly reflect the arrival times of event 2 at the actual positions of the detectors. 
Therefore, due to the ring degeneracy of a localization inferred from a phase difference, explained below Eq.\ (\ref{eq:tau_d1d2}), event 2 time delay phases will generically inherit a ring-like  structure that is typically bimodal. 
This bimodality can be separated into individual modes by  distinguishing the samples that come from above or below the plane of the detectors in their actual positions for event 2 (see App.\ \ref{app:breaking_bimodality} for details). By analyzing these two modes separately, we mitigate the non-Gaussianity of the inferred time delays for event 2.

In order to quantify the consistency of a set of posterior samples of parameters of event 1, $\theta_1$, with the same parameters of event 2, $\theta_2$, we focus on the probability density of their difference $\Delta \theta=\theta_1-\theta_2$ \cite{Raveri:2018wln}: 
\begin{equation}
    P(\Delta \theta) = \int P_{\theta_1}(\theta) P_{\theta_2}(\theta-\Delta \theta) \text{d}\theta\,,
\end{equation}
which is the convolution of the two posterior probabilities $P_{\theta_1}$ and $P_{\theta_1}$. Support for large values of
 $|\Delta\theta|$ indicates that the two parameter sets are not compatible.  

For any two posteriors, we can compute the distribution of $\Delta\theta$  numerically and set confidence intervals for consistency  \cite{Raveri:2021wfz}.  Moreover, 
if the two posteriors are approximately Gaussian, we can estimate consistency  very simply. 
The width $\sigma$ of the Gaussian $P(\Delta\theta)$ is determined by the sum of the covariances $C_1$ and $C_2$, so that the distance in units of $\sigma$ is given by \cite{Raveri:2018wln}
\begin{equation}\label{eq:gaussian_distance}
    D_{12}\equiv D(\theta_1,\theta_2) = \sqrt{\Delta \theta^T (C_1+C_2)^{-1} \Delta \theta}\,.
\end{equation}
Inconsistencies between data sets can be defined at different confidence levels (CL).  
Note that for non-Gaussian posteriors the Gaussian approximation in Eq.\ (\ref{eq:gaussian_distance}) will typically lead to overly conservative distances, as residual multi-modality will inflate the covariances leading to shorter distances. An example of this behavior for the time delay phases, which can be multimodal as described above, is presented in App.\ \ref{app:breaking_bimodality}.

For a given number of $\chi^2$-distributed parameters, we can relate the distance to the CL. For example, for 1/4/6 degrees of freedom, a 95\% CL corresponds to $2.0/3.1/3.5\sigma$, 99\% CL corresponds to $2.6/3.6/4.1\sigma$, and 99.9\% CL corresponds to $3.3/4.3/4.7\sigma$. 
If the uncertainties in a given parameter  make the data uninformative the effective degrees of freedom $N_{\rm eff}$ that are constrained by the data can be smaller than the total $N$.  In such cases the posterior has support across the whole  prior range and we can  quantify the effective number of degrees of freedom as
\begin{equation} \label{eq:neff}
    N_\text{eff} = \text{tr}\left[\left(C_\text{prior} + C_1 + C_
2\right)^{-1} C_\text{prior}\right]\,, 
\end{equation}
where $C_\text{prior}$ is the covariance of the priors. For example, for a bounded flat distribution in a single parameter the covariance is $C=(\theta_\text{max}-\theta_\text{min})^2/12$ \cite{Raveri:2018wln}. 
If the data is informative, $C_\text{prior}\gg C_1 + C_2$, then $N_\text{eff}\to N$, the true number, while in the opposite limit, $C_\text{prior}\ll C_1 + C_2$, one gets $N_\text{eff}\to0$.  
In practice given our 6 actual parameters, we compute $N_\text{eff}$ for the 3 detector phases only so that the total number of effective parameters is $N_{\rm tot}= N_\text{eff}+3$, effectively taking infinite prior covariances for the other parameters.  Because detector phases can differ under the lensing hypothesis by multiples of $\pi/2$, we set the detector phase priors $\phi_{d}$ for an informative measurement so that $C_{\rm prior} = \delta_{d_1,d_2}(\pi/2)^2/12 $ for $d_1,d_2 \in\mathrm{H,L,V}$. 

Since strong lensing introduces a constant phase shift of the detected phases, we compute the distance for all the possible phase shifts:
\begin{equation}
    D_{12}^{n}=D(\phi_1,\phi_2 + n\pi/2)\,,
\end{equation}
for $n=0,\pm1,\pm2$. 
A true lensed pair will have consistent parameters in both orderings and so we define a pair ordered distance that maximizes over the ordering:
\begin{equation}
    D^n_{J}=\text{max}(D_{12}^n,D_{21}^{-n})\,,
\end{equation}
where the relative phase shift and hence $n$ switches sign under a change in ordering. 
Conversely since each $n$ is a possible lensing outcome, we minimize the pair ordered distance over lensing types to obtain the final joint distance 
\begin{equation}
    D_J=\text{min}_n(D_J^n)\,.
\end{equation}
This $D_J$ will be our basic metric in the analysis. Note that by virtue of this formalism, we get directly the phase shift $n$ most consistent with the lensing hypothesis, which carries information about the image types and therefore the lens model. This is not possible with methods that compute the posterior overlaps that do not include phase information (see Sec.~\ref{sec:comparison}).

Finally, with the joint distance $D_J$ and the total number of effective parameters $N_\text{tot}$ we  compute the confidence level of consistency with the lensing hypothesis and produce a catalog of pairs that are consistent with lensing. 
In our analysis of simulated and real events in Secs. \ref{sec:simulatedevents} and \ref{sec:gwtc} we draw the line at $99\%$ CL but our same analysis could trivially output results at different levels, trading purity for completeness, though the probabilistic inference will depend on how Gaussian the tails of the parameter posterior are.

Since pairs that pass the distance consistency threshold may do so simply because the parameter errors are so large as to encompass lensing as a possibility, we can further rank them according to a statistic that measures how well the parameters are constrained. 
Because the errors in the parameters for a given event scale  inversely with the SNR,    
one possibility is to rank pairs in terms of expected scaling of total errors with the network SNR
\begin{equation} \label{eq:R}
R= \sqrt{\rho_{\rm ntw,1}^{-2} + \rho_{\rm ntw,2}^{-2}}
\end{equation}
where
\begin{equation}
\rho_{\rm ntw}^2 = \rho_\mathrm{H}^2 + \rho_\mathrm{L}^2 + \rho_\mathrm{V}^2
\end{equation}
is the median ``optimal SNR'' across the posterior parameter distribution (see Eq.~\ref{eq:optimal_snr}), combining those of each detector.  
The drawback is that there can be events with high SNR and poor parameter errors when compared relative to different events with very different binary parameters. 
In fact, as discussed above when the 22-mode is the dominant detection, all detector phase errors should scale with $R$ with the same  order unity coefficient $(\sim 3)$  so that a parameter error based ranking can be constructed using just the parameter volume of the three detector phases $\theta \in \{\phiH,\phiL,\phiV\}$ 
\begin{equation}
    V_{\theta} = \sqrt{\mathrm{det}(C_\theta)}\,.
\end{equation}
For a pair we take the sum over each volume
\begin{equation}
    V_J = V_1+V_2\,.
\end{equation}
Unlike for the parameter distance and $N_{\rm eff}$, this covariance and the associated volume is computed in the reference frame of individual events with the detectors in their true orientations so as to better reflect the scaling with SNR of all of the derived parameters.  

In our default analysis we compute the parameter distance and consistency of the six phases: $\{\phiH,\phiL,\phiV,\Delta\phi,\tauHL,\tauLV\}$. This can be extended to include any additional parameter, properly including their correlations. In particular, in the future, the polarization amplitude ratio $r(f)$ could be an additional discriminator that measures spin or precession consistency, though it is currently not independently well measured. 

As we will show in the next sections, since the phase difference is the more constraining parameter at the moment, when  constructing a candidate lensing catalog we recommend employing both $D_J$ and $\Delta\phi_f$, rejecting those pairs inconsistent with either at the chosen CL.  

Our method is implemented in a public code: \gwphase\footnote{\gwphaselink}, which includes both the post processing tools to rapidly obtain the detector phases as well as the tension statistics that are relevant for lensing searches.

\begin{figure*}[t!]
\centering
\includegraphics[width = 0.7\textwidth,valign=t]{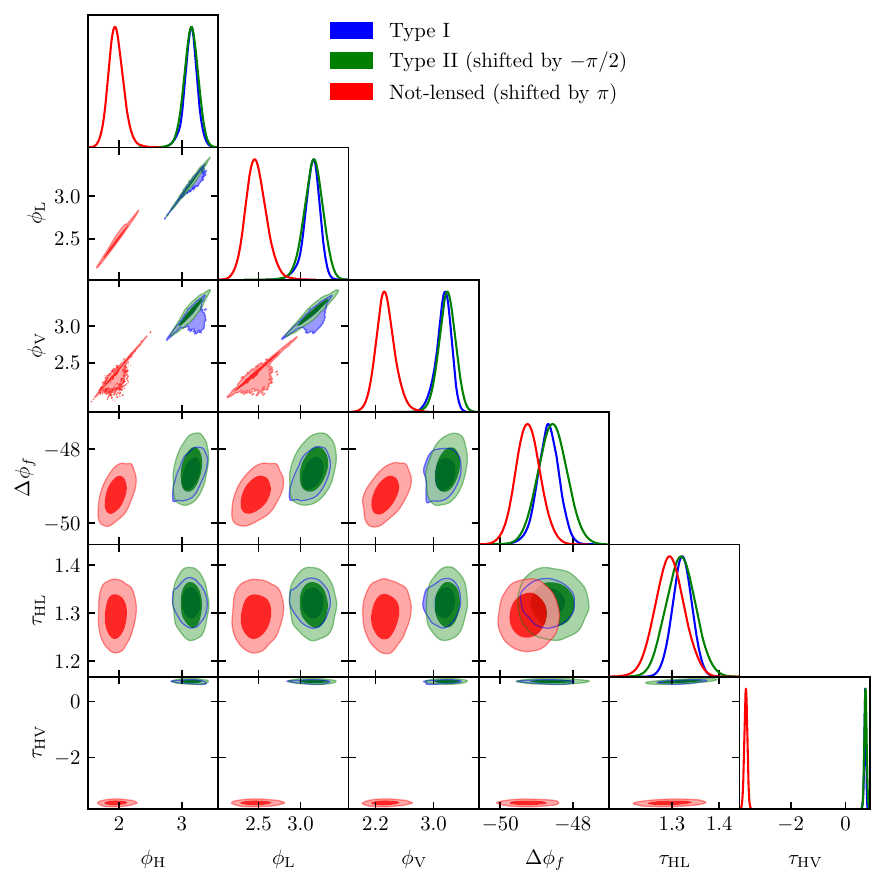}
\caption{
Lensing consistency test for a simulated pair of lensed events (type I and type II images) and a not-lensed event. The reconstructed phases at each detector ($\phi_d$) and the time delay phases ($\tau_{d_1d_2}$) are computed at 40Hz, while the orbital phase evolution ($\Delta\phi_f$) is between 20 and 100Hz. 
We are comparing the consistency of both the type I image and the not-lensed event to the type II image. 
For the three injections the posteriors correspond to the high signal-to-noise, 3-detector configuration (specifically Injections 10-12 described in App.\ \ref{app:injections}). 
In this six dimensional phase space, the lensed pair is consistent within a distance of $0.6$, while the not-lensed pairs are rejected with distances $>70$. 
Contours are drawn at 68 and $95\%$ credible intervals.}
\label{fig:lensing_consistency}
\end{figure*}

\section{Analysis of simulated events}
\label{sec:simulatedevents}

In order to test our method, we simulate a set of lensed and not-lensed GW detections, which we refer to as injections. 
Our set of injections contains a reference GW event with zero phase shift  denoted ``Type I", an image of this event (demagnified and phase shifted by $\pi/2$) denoted ``Type 2", and a third unrelated event which is not-lensed, denoted ``Not-lensed". 
This accounts for 1 lensed pair (type I \& II) and 2 not-lensed pairs (type I/II \& Not-lensed). 
Within each set we tune their luminosity distance so that there is a low-SNR configuration ($\rho_\text{ntw}\sim12-16$) and a high-SNR configuration ($\rho_\text{ntw}\sim22-30$) across the detector network.   
We study a LIGO-only two detector network (HL) and a LIGO-Virgo three detector network (HLV). 
We make these choices to represent different observing scenarios.  
In all the cases we perform the standard parameter estimation that is applied to real data in LVK catalogs using \texttt{bilby} \cite{Ashton:2018jfp} and the projected sensitivities of LIGO and Virgo for the fourth observing run, see App.\ \ref{app:injections} for details. 
Our goal is to test the robustness of the method for situations in which it is easy to confuse a pair of unlensed events as being lensed due to the similarity of their parameters.

We first consider the case of a ``vanilla" binary black hole with masses comparable to GW150914, i.e.\ $\mathcal{M}_c\sim30M_\odot$. 
In order to test the method in a situation in which standard posterior overlap analysis could lead to false alarms, we consider the case in which the non-lensed injection has the same intrinsic parameters as the lensed injection. We only change the phase, polarization angle and sky position. The sky position is chosen so that for a LIGO-only detector network the sky maps of the lensed and not-lensed events overlap, i.e.\ on the degeneracy ring of the time delay $\tauHL$ (see Fig.\ \ref{fig:time_delay_contours}). 
Details of these injections are given in Tab.\ \ref{table:injections_parameters}, Injections 1-12.

To get a sense of the method, we first consider the most optimistic scenario with high SNR and an HLV network. We present the posterior distributions for the different phases of the three injection in Fig.\ \ref{fig:lensing_consistency}. 
The lensed pair, Type I - Type II, displays agreement when shifting the detector phases by $-\pi/2$, while it is clear by eye that the not-lensed event shows inconsistency with them. 
To quantify the (in)consistency of the pairs, we compute the distance $D_J$ of the simulated lensed and not-lensed pairs.  
For the lensed pair, the minimum distance in the six dimensional phase space is $0.6$ when shifting the type II image phases by $-\pi/2$. Any other phase shift, $0,\pi/2,\pm\pi$, would lead to distances $>11$. 
On the other hand, for the not-lensed pairs, all possible phase shift configurations lead to  disagreements with $D_J>70$. 
This large distance is dominated by the time delay phase $\tauHV$ that alone rejects the lensing hypothesis with a distance of $67$. 
However, the three detector phases alone would also strongly reject lensing, with a three dimensional distance $D_{\phi_d}>15$, thanks to the well constrained direction $\phiH-\phiL$. 
On the other hand, in this case where we have intentionally set the mass to be the same the orbital phase evolution $\Delta\phi_f$ alone would agree within a one dimensional distance of $D_{\Delta \phi_f} \sim1$, consistent with noise.  
In any of the pairs, all three detector phases are informative, with a total effective number of detector phases $N_{\rm eff}=2.7$
and a total number of effective parameters $N_{\rm tot}=5.7$.  For this number of parameters both $D_J>70$ and $D_{\phi_d}>15$ indicate an entirely negligible probability of being a lensed pair.

Moving forward, our goal is to test how well we can reject the non-lensed hypothesis for pairs that share many common parameters in different configurations. 
For this first set of 12 injections we study all the possible low/high-SNR and HL/HLV configurations for a total of $12\times 11/2 = 66$ possible pairs. 
Each injection has a different noise realization and their parameters are summarized in App.\ \ref{app:injections}.  
We find that the not-lensed pairs are \emph{always} rejected with distances $D_J>3$ in the six dimensional phase space or, equivalently, $>94\%$ CL when taking into account the effective number of informative degrees of freedom. 
Only 2 of the 32 not-lensed pairs are consistent with lensing at 99\% CL even in this situation in which most of the parameters of the simulated GWs are the same.   
For injections with high-SNR or HLV configurations (9 out 12), non-lensed pairs are \emph{always} rejected with distances larger than $5$. 
On the other hand, we find all lensed pairs to be consistent within a distance of $2.5$, where recall that the median expectation for $N_{\rm tot}=6$ is $\sim 2.3$.  

\begin{figure}[t!]
\centering
\includegraphics[width = \columnwidth,valign=t]{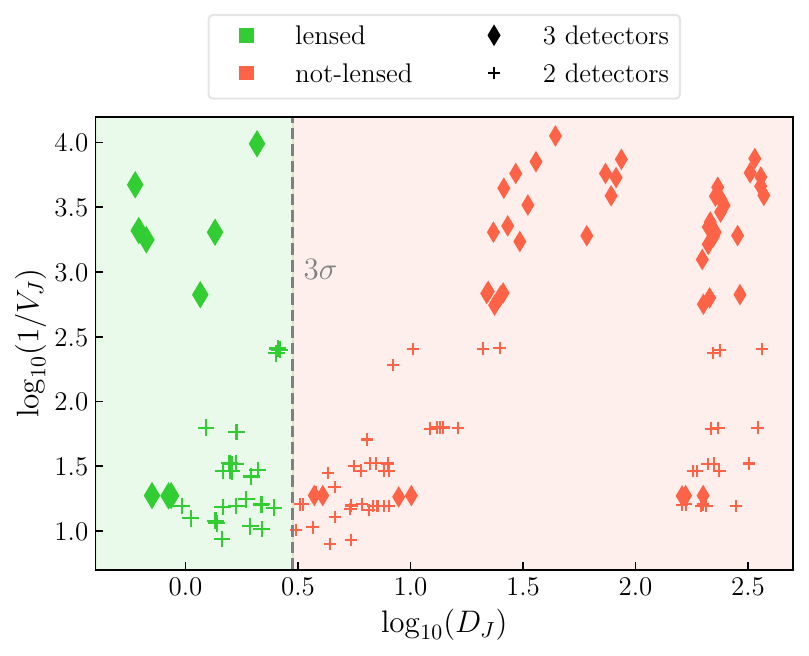}
\caption{Analysis of simulated gravitational wave events. We compare the distance ($D_J$) and  phase volume ($V_J$) statistics for all 136 pairs from 17 injections. Lensed and non-lensed pairs are indicated with green and red colors respectively.  Point shapes  indicate pairs in which in both events there were HL ($+$)  or  HLV-detectors ($\blacklozenge$). 
The vertical dashed line delineates  3$\sigma$ in distance ($D_J=3$) and note that all lensed pairs lie below this threshold (shaded green) and all non-lensed pairs above (shaded red). 
The volume serves to quantify how well constrained the detector phases are compared to their prior volume and can be used to rank candidates for followup analyses, from larger to smaller $1/V_J$. }
\label{fig:distance_injections}
\end{figure}

The distribution of distances for all simulated pairs is presented in Fig.\ \ref{fig:distance_injections}. 
There we also plot the detector phase volume in the vertical axis. 
The distance itself is the one determining the consistency with the lensing hypothesis.  
$V_J$, on the other hand, provides the ranking of the events that pass the lensing consistency test for follow-up analyses. In particular, we clearly see that HLV pairs have smaller volumes as expected.   

To understand how the masses of the binary play a role in the method, we then study a lower mass binary with $\mathcal{M}_c\sim 12M_\odot$ (corresponding to Injections 13-15 in Tab.\ \ref{table:injections_parameters}). 
The relation between the three injections is the same as before, but in order to enlarge the cases tested we choose a different inclination and sky position to the previous case for injection 3 and change the arrival times in all cases. This defines the ``low mass" version of the Type I, Type II and Not-lensed injections.  
In this case we only consider the low-SNR, HLV configuration. Therefore, when putting this together with the previous injections we analyze a total of 15 injections and 105 possible pairs. 
We find similar results as before, rejecting the not-lensed pairs with distances $>3$, see Fig.\ \ref{fig:distance_injections}. 
No additional not-lensed pair is consistent with lensing at 99\% CL, reflecting the fact that the original set was designed to be difficult to distinguish, and so the false-alarm rate goes down to 2 out of 70 of the injection pairs.  
Moreover, this new set of injections allow us to compare with the previous injections to see how two events with all equal parameters but masses and arrival times could be distinguished. Thanks to incorporating the phase difference $\Delta\phi_f$, events with different masses are rejected at higher significance. Moreover, we find that the phase difference can be a more stringent discriminator than the chirp mass. 
In this case {$D_{\Delta\phi_f} \sim 10 D_{\mathcal{M}_c} \sim 100$}, though as we shall see in the next section a more typical number for $D_{\Delta\phi_f}/D_{\mathcal{M}_c}$ is $2-3$ for distances that are more consistent with the lensing hypothesis. 

Finally, we look at the effect of inclination. We do so by simulating high-SNR HLV triggers with the same parameters of Type I and II, but with a different inclination, $\theta_{JN}=2.3$ compared to $\theta_{JN}=0.8$, cf.\ Injections 10,11 to 16,17  in Table \ref{table:injections_parameters} for details. 
We find that for all pairs  
with these different inclinations, the lensing hypothesis is rejected through the phases at the detectors. This implies that $\phi_d$ encodes the inclination information,  or circular polarization ratio $r$, as expected. 
We also look for the agreement in the polarization ratio $r$ alone. 
We find that only HLV-injections are able to discern between the two cases due to parameter degeneracies. 
In particular, the inclination will be degenerate with other polarization state parameters in generic precessing systems. An example of this behavior is presented in Fig.\ \ref{fig:polarization_evolution} in App.\ \ref{sec:phase_new_freq}. 

Altogether, for the 17 injections and 136 pairs, including both HL and HLV configurations,  we find that the detector phases can carry relevant information. When computing the number of effective degrees of freedom, we find that all pairs have at least one effective phase, and $>40\%$ have 2 effective phases, with $>25\%$ having $N_\text{eff}>2.5$ (a compilation of the detector phases posteriors is presented in App.\ \ref{app:injections}). 
However, given our specific choices for the injections, this does not represent a full characterization of an astrophysical population and observing run. For example, in this catalog of simulated events, $\sim40\%$ of pairs are HLV detections. 
A detailed study with a fair selection  of simulated not-lensed events from an astrophysical population for different observing scenarios is left for future work, though as we shall see next, the known real events provide a proxy.

\section{Analysis of real events} \label{sec:gwtc}

Having validated our method with simulations in the previous section, we proceed to analyze the latest LVK catalog, GWTC-3 \cite{LIGOScientific:2021djp}.\footnote{It is to be noted that GWTC-3 includes the catalog GWTC-2.1 \cite{LIGOScientific:2021usb} that reanalyzed the events in the first two observing run with the newest waveforms used during the third observing run.} 
We focus on binary black holes with a high probability of having astrophysical origin, $p_\text{astro}>0.8$, and a network SNR $>8$, for a total of 67 events or 2211 pairs.

\begin{figure}[t!]
\centering
\includegraphics[width = \columnwidth,valign=t]{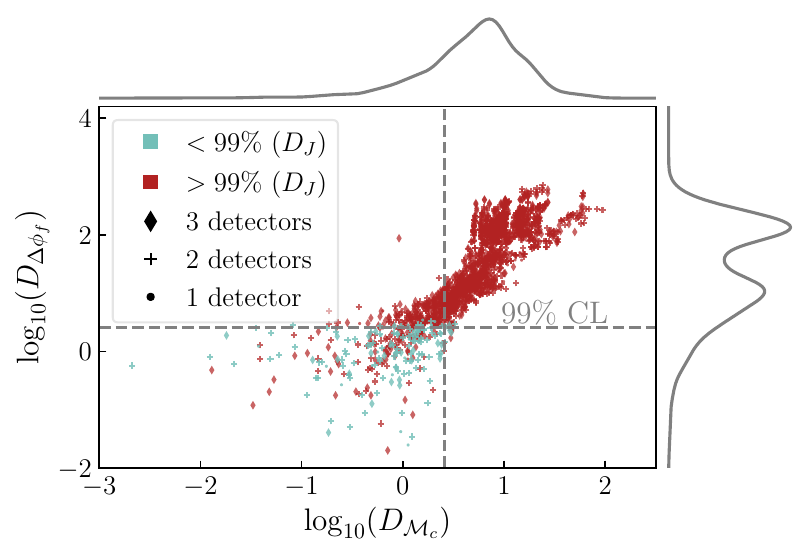}
\caption{Comparison of the one dimensional distance in the derived orbital phase evolution $\Delta\phi_f$ and the detector frame chirp mass $\mathcal M_c$ for real gravitational wave events. $\Delta\phi_f$ efficiently rejects 89\% of the pairs as being not-lensed at 99\% CL, while the same threshold on $\mathcal M_c$  more than doubles the number of the pairs. Colors denote the 99\% CL cut on $D_J$ alone and note that some pairs are ruled out by $D_{\Delta\phi_f}$ but not $D_{J}$ (see text for discussion). Point shapes indicate pairs in which both events had \emph{at least} 3 detectors ($\blacklozenge$), 2 detectors ($+$) or 1 detector ($\bullet$) online.} 
\label{fig:1D_distance_gwtc}
\end{figure}

With the current catalog, where the chirp masses are distributed over a wide range of values and the Virgo detections are low SNR at best, we expect the main component of the distance discriminator to be the 
phase difference between frequencies $\Delta \phi_f$. 
As previously discussed, $\Delta\phi_f$ is a better constrained parameter than the detector-frame chirp mass, leading to distances which are typically at least a factor of 2-3 larger as shown in Fig.\ \ref{fig:1D_distance_gwtc}. 
This leads to a significant advantage when testing the lensing hypothesis.
If we compute the 1D distance consistency $D_{\Delta\phi_f}$ determined by $\Delta\phi_f$ alone, we find that 
only 11\% of the pairs are consistent 
with the lensing hypothesis at 99\% CL.  
This is to be compared with the $\mathcal{M}_c$ consistency alone where more than double that number 23\% are consistent with lensing at 99\% CL.

\begin{figure}[t!]
\centering
\includegraphics[width = \columnwidth,valign=t]{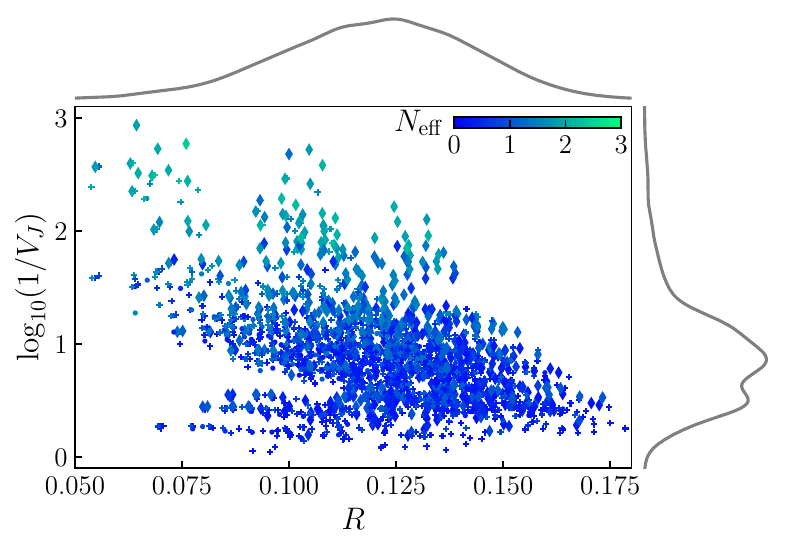}
\caption{ 
Detector phase volume $V_J$ as a function of $R$ for real gravitational wave events. $R$ weights the inverse network SNR of the pair, see Eq.~(\ref{eq:R}). We color the pairs by their number of effective detector phases $N_{\rm eff}$.  
Smaller $R$ tends to correspond to larger $1/V_J$ and a higher $N_{\rm eff}$ albeit with large scatter, mainly associated with the poorer phase measurements for high chirp mass events. 
Point shapes indicate the number of detectors as in Fig.\ \ref{fig:1D_distance_gwtc}.
} 
\label{fig:vol_vs_R}
\end{figure}

While the joint distance $D_J$ automatically accounts for $D_{\Delta\phi_f}$, the detector phase differences are only significant in a fraction of the pairs in the current catalog.
Quantitatively,  only 31\% of the pairs have more than one effectively constrained detector phase at 40Hz.  
In other words, for most of the pairs, the detector phases have errors larger than $\pi/2$, a precision that is necessary to distinguish between all the possible lensing phase shifts (and that is a factor of 4 smaller than the naive $2\pi$ prior range of the phases, requiring a factor of 4 higher SNR to be informative). 
The number of pairs with $N_\text{eff}>1$ increases to ${45\%}$ when focusing on pairs with three detectors online. 
As shown in Fig.\ \ref{fig:vol_vs_R}, pairs with a low number of effective phases are those with large volumes and $R$ statistic (see Eq.~\ref{eq:R}). 
Note though that the scatter around this trend is large and that the pairs in Fig.\ \ref{fig:vol_vs_R} with small $R$ but large volume are associated to events with high chirp mass.

The large scatter to higher $V_J$ also implies that there can be cases where lensing is excluded by $D_{\Delta\phi_f}$ but marginally allowed by $D_J$. 
Within the current catalog and a 99\% CL threshold, this occurs in 16 pairs.  Recall that we compute the $p$-value for $D_J$ using the $N_\text{eff}$, see Eq.\ (\ref{eq:neff}).  
In order to avoid cases where a strong rejection in the one dimensional space is diluted by poor and potentially non-Gaussian measurements in the higher dimensional phase space, 
we require both $D_{\Delta\phi_f}$ and $D_{J}$ to pass the 99\% CL\ consistency test in the current catalog.
This brings our total to 131  out of 2211 pairs or $\sim 6\%$ that are consistent with lensing.

\begin{figure}[t!]
\centering
\includegraphics[width = \columnwidth,valign=t]{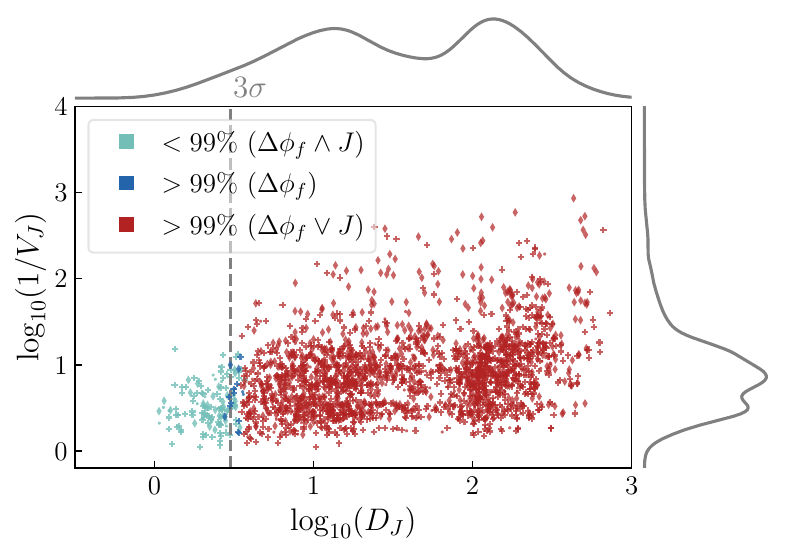}
\caption{Analysis of real gravitational wave events. We compare the distance ($D_J$) and detector phase volume ($V_J$) statistics for 67 binary black holes in GWTC3 with probability of having astrophysical origin larger than 0.8 and network SNR $>8$. 
We use different symbols to indicate the number of detectors online and different colors to show the confidence level of agreement with the lensing hypothesis for both the joint phase distance and phase evolution ($\Delta \phi_f \land J$), for only phase evolution ($\Delta \phi_f$) or for neither. 94\% of pairs are rejected at 99\% CL. 
Point shapes indicate the number of detectors as in Fig.\ \ref{fig:1D_distance_gwtc}.}
\label{fig:distance_gwtc}
\end{figure}

We display these cuts and the final lensing catalog in Fig.~\ref{fig:distance_gwtc}.
As expected, 3 detector (HLV) events with higher SNR display smaller volumes and reject the lensing hypothesis more strongly.
Notice also that some of the structure in the distribution of distances and volumes in Fig.\ \ref{fig:distance_gwtc} is inherited 
from the astrophysical population of chirp masses discussed in 
Fig.~\ref{fig:vol_vs_R}.

For each pair in the final lensing catalog, our method provides  information about the image types through the phase shift which gives the minimum distance. 
The pair GW170104--GW170814 stands out as the most interesting candidate given its small phase volume, $\log_{10}(1/V_J)=1.2$, and distance, $D_J=1.3$, for a phase shift of $\pm\pi$. This would correspond to type I-type III image pairs. 
This pair was identified early on by \cite{Dai:2020tpj,Liu:2020par}, and remains the pair with highest coherence ratio \cite{LIGOScientific:2023bwz,Janquart:2023osz} (see also Sec.~\ref{sec:comparison}). When including astrophysical priors about the lens model and population of lenses, the lensing hypothesis is however disfavored under those model assumptions \cite{LIGOScientific:2021izm}. 

\section{Comparison with other methods} \label{sec:comparison}

GW strongly lensed candidate pairs can be ranked by the amount of overlap of their parameters. 
This posterior overlap method \cite{Haris:2018vmn} uses
\begin{equation}
    \BLU = \int \mathrm{d}\Theta \frac{p(\Theta|\mathrm{event}_1)p(\Theta|\mathrm{event}_2)}{p(\Theta)}
\end{equation}
as its basic statistic where $\Theta$ are the set of parameters over which overlaps are computed, $p(\Theta|\mathrm{event})$ are their posterior distributions and $p(\Theta)$ their prior distribution. 
Previous analyses \cite{Hannuksela:2019kle,LIGOScientific:2021izm,LIGOScientific:2023bwz,Janquart:2023mvf} have computed overlaps over 8 parameters: detector-frame masses, dimensionless spin magnitudes, the cosine of spin tilt angles, the cosine of orbital inclination $\theta_{JN}$ and sky position. 
The $\BLU$ statistic is therefore a ratio of marginalized posteriors that  is agnostic to astrophysical assumptions. 
The $\BLU$ is however sensitive to the prior range of its parameters, in particular the detector frame masses. 
Moreover, being a statistic that it is not normalized or calibrated intrinsically, the interpretation of a $\BLU$ is subject to prior knowledge of the expected values for a population of lensed and not-lensed sources. Aided by background studies, one can then translate a $\BLU$ into an odds ratio or $p$-value \cite{LIGOScientific:2021izm,LIGOScientific:2023bwz}. 
Since we want to compare the posterior overlap method with our phase consistency approach that it is agnostic to astrophysical modeling or assumptions about the likelihood of the lensed and not-lensed hypotheses, we therefore restrict to $\BLU$ and interpreting it as a Bayes factor.   Correspondingly, for illustration purposes, we  assume that only pairs with $\BLU<0.1$ reject the lensing hypothesis. 

In the past $\BLU$ has been used to select interesting candidates to follow with joint parameter estimation. In the first half of the third observing run 19 pairs of super-threshold events with $\BLU>50$ were followed up \cite{LIGOScientific:2021izm}, although 4 of them involved a GW event (GW$190424\_180648$) that was later downgraded to sub-threshold \cite{LIGOScientific:2021usb}. 
In \cite{LIGOScientific:2021izm}, the O2 pair GW170104-GW170814 was also analyzed in joint PE. 
In the second half of O3 pairs with a false positive probability (FPP) of $<1\%$ were further analyzed \cite{LIGOScientific:2023bwz}, with the FPP computed after a large injection campaign. In the second half of O3, events were also filtered using the machine learning code LensID \cite{Goyal:2021hxv} and the rapid joint parameter estimation version of Golum \cite{Janquart:2023osz}. This accounted for another 14 joint parameter estimation pairs. 
In total there are 30 pairs of super-threshold events that we can compare with.
For all these pairs, the joint parameter estimation provides the coherence ratio $\CLU$, defined as the ratio of the lensed and not-lensed evidences, and it is (mostly) agnostic to astrophysical priors on the population of sources and lenses \cite{LIGOScientific:2021izm}. 
LVK analyses \cite{LIGOScientific:2021izm,LIGOScientific:2023bwz} also computed the Bayes factor that accounts for the final likelihood of lensing taking into account reasonable astrophysical expectations. 

\begin{figure}[t!]
\centering
\includegraphics[width = \columnwidth,valign=t]{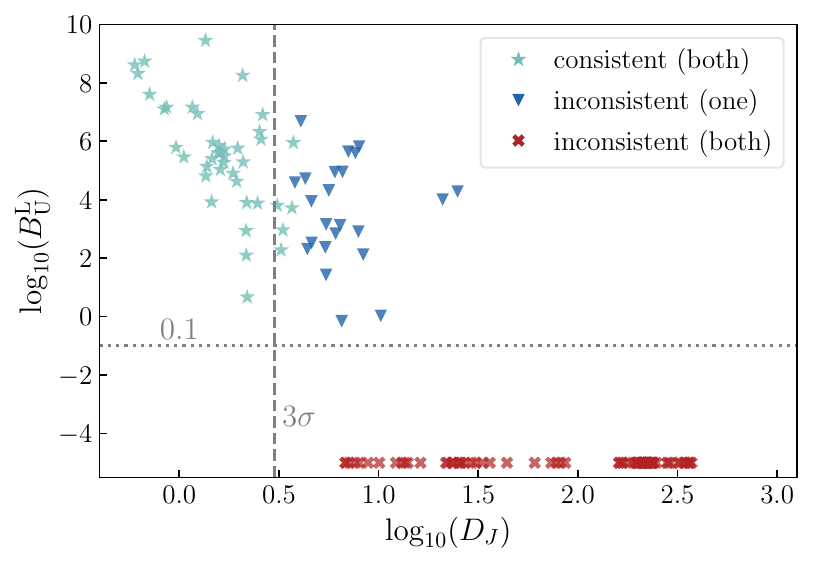}
\caption{Comparison between the phase consistency and the posterior overlap methods for simulated gravitational wave events. Phase consistency is determined by the distance $D_J$, while posterior overlap is quantified by the $\BLU$ statistic. We follow the same plotting conventions of Fig.\ \ref{fig:distance_injections}. 
The minimum $\BLU$ is set to $10^{-5}$.}
\label{fig:phase_vs_overlap_injections}
\end{figure}

Following \cite{Hannuksela:2019kle,LIGOScientific:2021izm,LIGOScientific:2023bwz}, we compute the posterior overlap of a given pair approximating the posterior distributions with a kernel density estimator (KDE) for the first seven parameters and computing the sky overlap independently. Prior volumes are specified in App. \ref{app:gwtc}. 
Importantly, the sky overlaps in this approach are not normalized. 
Alternative sky map overlap statistics have been studied in \cite{Wong:2021lxf}. 
We note that the numerical calculation of the six-dimensional KDE is significantly more demanding than the distance calculation, which is typically 1000 times faster. 

We begin by analyzing our injection set. Fig.\ \ref{fig:phase_vs_overlap_injections} presents the distance statistic $D_J$ versus the posterior overlap $\BLU$ for all possible pairs. 
Noticeably most lensed pairs display a large $\BLU$. 
However, there are also many not-lensed pairs with similar overlap that are rejected by their phase consistency. Moreover there are no pairs with $\BLU<0.1$ and $D_J>3$ which would represent a more efficient rejection via overlap.  This demonstrates the potential of our method to better reduce the number of false alarms. 
Because of not being normalized, the overlap statistics show a tendency to predict HLV-pairs with higher $\BLU$, regardless of lensing. 

\begin{figure}[t!]
\centering
\includegraphics[width = \columnwidth,valign=t]{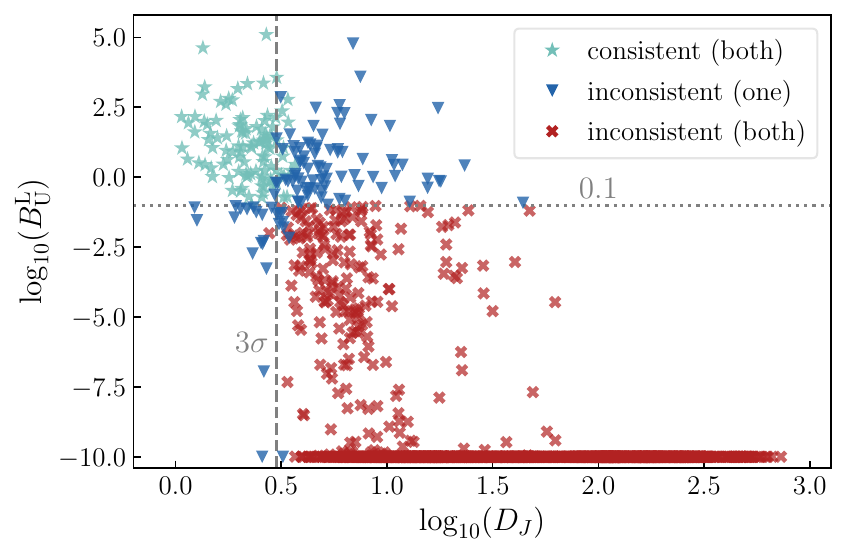}
\caption{Comparison between the phase consistency and the posterior overlap methods for real gravitational wave events. We follow the same plotting conventions of Fig.\ \ref{fig:distance_gwtc}. 
The minimum $\BLU$ is set to $10^{-10}$.}
\label{fig:phase_vs_overlap_gwtc}
\end{figure}

We then analyze the catalog of real binary black holes in Fig.\ \ref{fig:phase_vs_overlap_gwtc}. 
We find that with the posterior overlap statistic $9\%$ of pairs are consistent with the lensing hypothesis as determined by $\log_{10}\BLU > -1$. This is to be compared to the $6\%$ found with the phase consistency at 99\% CL. 
The rejection of the lensing hypothesis in the posterior overlap analysis is mainly triggered by non-overlapping sky maps, with 30\% of pairs having no sky-overlap. 
This is due mostly to the majority of events in the catalog being relatively heavy (due to selection effects) leading to larger errors in their masses, and agrees with previous expectations \cite{Caliskan:2022wbh}. 
When excluding the sky localization information, 18\% of pairs are consistent with lensing.  

In terms of the purity of the $\BLU>0.1$ pair catalog, the phase consistency method disfavors (at 99\% CL) 44\% of candidate pairs, for a total of 82.  In Fig.~\ref{fig:phase_vs_overlap_gwtc}, these are the upper right blue points. 
From them, some exhibit strong consistency with the lensing hypothesis through their overlaps, $\BLU>100$, but are rejected by the phase consistency at high confidence. This is because in the posterior overlap method a single parameter can outweigh the rest, while the phase consistency accounts more equally from the consistency in all the phases. 
This demonstrates the complementarity and advantages of our new method compared to existing analyses. 

Conversely, in terms of the purity of the phase consistency catalog, 19\% of pairs consistent with lensing by their phases have $\BLU<0.1$.
In Fig.~\ref{fig:phase_vs_overlap_gwtc}, these are the 25 lower left blue points, i.e.~less than a third of the false alarms of $\BLU$. 
These are cases in which the Gaussian distance estimator provides an overly conservative agreement for an intrinsically non-Gaussian posterior, especially in the time delay phases of two detector events where the degeneracy forms rings (see App.\ \ref{app:breaking_bimodality}).  In this case, spot checking our phase-based pair catalog for sky overlap consistency can be helpful.  More generally, one can automate tests for significantly non-Gaussian distributions such as a difference between the mean and mode.  For such cases, 
a non-Gaussian tension estimation for phase consistency can be efficiently achieved using machine learning methods \cite{Raveri:2018wln}. We leave the implementation of such tools for future work. 

\begin{figure}[t!]
\centering
\includegraphics[width = \columnwidth,valign=t]{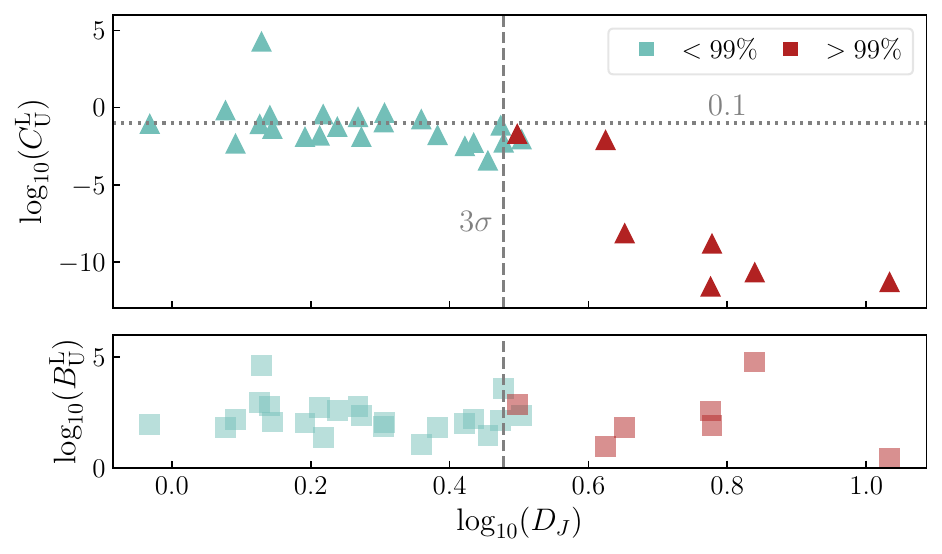}
\caption{ 
Comparison between the phase consistency distance ($D_J$), the joint parameter estimation coherence ratio ($\CLU$) and the posterior overlap statistic ($\BLU$) for real gravitational wave events. The joint parameter estimation results are taken from \cite{LIGOScientific:2021izm,LIGOScientific:2023bwz}. }
\label{fig:phase_vs_clu_gwtc}
\end{figure}

Finally, we compare our phase consistency distance statistic $D_J$ with the joint parameter estimation coherence ratio $\CLU$ and the posterior overlap $\BLU$ for the 30 pairs in which this information is available. 
As shown in Fig.\ \ref{fig:phase_vs_clu_gwtc} all pairs display a large $\BLU$ since this was the criteria for joint parameter estimation follow up. We can also see that only  GW170104-GW170814 favors lensing with $\log_{10}\CLU>0$. In fact, many of them are rejecting the lensing hypothesis $\log_{10}\CLU<-1$. 
Noticeably, our distance statistic is able to reject 6 pairs at 99\% CL with large $\BLU$ but small $\CLU$. 
The anticorrelation between $D_J>3$ and $\CLU$ demonstrates the benefit of incorporating the phase consistency in future strong lensing searches. 
Our method also correctly identified the pair GW170104-GW170814 as the most significant given the consistent distance and small volume, as discussed in the previous section. 

\section{Current and future strong lensing searches}
\label{sec:strong_lensing_searches}

\begin{figure}[t!]
\centering
\includegraphics[width = \columnwidth,valign=t]{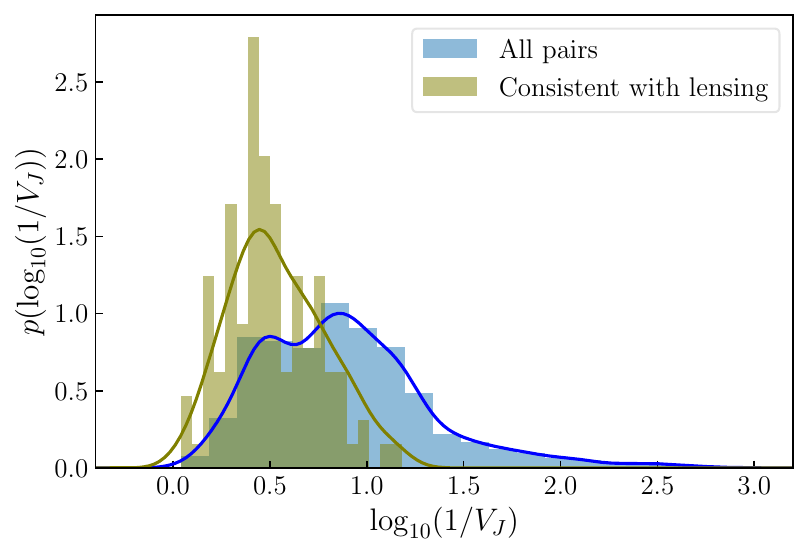}
\caption{ 
Distribution of detector phase volume for all the pairs analyzed in GWTC3, and those consistent with the lensing hypothesis. Consistency with lensing is defined at 99\% CL as in the main text. This represents $6\%$ of the pairs.
}
\label{fig:distribution_vol_phase}
\end{figure}

Our analysis of real GW events in the first three observing runs together with the simulations of lensed and not-lensed events with the sensitivity of the fourth observing run allow us to draw some expectations for the upcoming and future multiple image searches. 
As we have seen in the previous sections, our ability to test the lensing hypothesis depends on how well we can reconstruct the different phases.  
Past detector sensitivities were such that the distance measuring the (in)consistency of a given pair with the lensing hypothesis was dominated by the orbital phase evolution information, as the detector phases were typically not well enough constrained to distinguish between the possible lensing phase shifts (see Fig.\ \ref{fig:detector_phases_gwtc} in App.\ \ref{app:gwtc} for the actual posterior distributions of $\phi_d$). 
In fact, when looking at the distribution of detector phase volumes, see Fig.\ \ref{fig:distribution_vol_phase}, one can see that pairs consistent with the lensing hypothesis are skewed towards larger volumes, making ranking by $V_J$ even more important. In other words, most of the pair catalog is composed of events with poor parameter constraints and/or low SNR.

The situation with our simulated events shows more promise for the future. During the fourth observing run there should be more events where the phases are better constrained and provide a better discriminant for the lensing hypothesis. Still, we observed large difference in the constraining power of our optimistic, high SNR simulated events to our more common, low SNR injections and in any given observing run low SNR events will always be more numerous than high SNR events. 

In general, for the detector phases to be informative we need to distinguish between the possible $\pi/2$ phase shifts within the pair. 
Thus, the effective $\sigma$ from the lensing prior is $\pi/2\sqrt{12}$. From data SNR per detector is $3\sqrt{2}/\rho_d$, so nominally for the detector phases to be informative it is necessary that $\rho_d > 12\sqrt{6}/\pi \sim 10$ or $R<\pi/12\sqrt{6}\sim0.1$ for a two detector network. 
Within the real events that we analyzed in GWTC3, that is true for only about $20\%$ of pairs, see Fig.\ \ref{fig:vol_vs_R}. 
The distribution of SNRs for a given detector network is (quasi-)universal \cite{Schutz:2011tw}.\footnote{The strict universality holds in an Euclidean universe where sources are uniformly distributed in distance. This is a good approximation for low-redshift observations.}  
It follows $p(\rho_\text{ntw})=3\rho_\text{th}^3/\rho_\text{ntw}^4$, where $\rho_\text{th}$ is the threshold for detection. Therefore, it is to be anticipated that for a fixed network, the majority of events will be of low significance close to the threshold. Still, future observing runs will improve their sensitivity, making distant events with higher chances of experiencing lensing better measured and easier to identify. 
Within our current data, if we increase the network SNR cut to 12, the catalog size reduces to 36 events (630 pairs) and the fraction of pairs consistent with lensing also reduces to $3\%$, or 18 pairs. 
Given the current computational cost of joint parameter estimation analyses, reducing the number of pairs to follow up in a ``golden catalog" of lensed candidates would be advantageous and with our technique can be achieved through cuts in $V_J$. 
As demonstrated in Fig.\ \ref{fig:phase_vs_clu_gwtc}, our distance measure correlates nicely with the joint parameter estimation coherence ratio and can be used to reduce the candidate list compared to other methods. 

Besides the GW information, electromagnetic follow ups of strongly lensed candidates will be essential to conclude the lensed origin of a set of GW events. This can be achieved in different fashions. Targeted follow ups and archival searches can look for lensing signatures of the host galaxy or other galaxies in the line of sight of the joint sky localization of the pair \cite{Hannuksela:2020xor,Wempe:2022zlk}. Similarly, one may search for other lensed transients that might be associated with the GW signals \cite{Smith:2022vbp}. 
In any case, information about the lens model is key to achieve an efficient cross-correlation. 
In that respect, our method is also advantageous as it provides a fast lensing consistency check with additional information about the image types through the preferred lensing phase shift. 
Image type information together with the time delay and magnification ratios can be used to model the lens. 

\section{Conclusions} \label{sec:conclusions}

The detection of lensed gravitational waves remains elusive in current datasets. The key signature for GWs traveling through a massive lens would be repeated copies of the emitted signal, arriving at different times, sharing the same frequency evolution and polarization state, but with different amplitudes and specific differences in absolute phases. The quest of discovering strongly lensed GWs thus entails identifying these repeated chirps within a large catalog of not-lensed events. The crux is that many not-lensed pairs of events may look alike, simply because their parameters are not well enough constrained to discern the difference.

We have developed a new method to efficiently reject event pairs that are inconsistent with the lensing hypothesis and construct a catalog of candidates for further study. Our method's main ingredients are: 1) reconstructing the best measured GW parameters and 2) determining the consistency of the events rather than account for the degree of their parameter overlap.  This design allow us to efficiently produce a highly complete catalog that is also more pure than previous methods.  
Efficiency comes from the fact that we post-process the full parameter estimation posterior samples to obtain the phases that would have been measured by each detector. We demonstrate that at 40Hz such phases are constrained within a fraction of a radian and that its change through the frequency spectrum precisely quantifies the orbital phase evolution with better relative errors than the chirp mass. Moreover, we show that the phases associated to the arrival time differences of the signal at the detectors are also well constrained and reduce the degeneracies of typical sky localizations.  

The degree of completeness vs.\ purity is controlled by computing the distance between two events in the multi-dimensional space of their phases. Because the phases are better constrained, their distributions are more Gaussianly distributed than the original waveform parameters and so we apply a Gaussian approximation to the distance whose evaluation is computationally trivial. 
From the pairs consistent with lensing, we rank them by their phase volume so as to de-emphasize the remaining candidate pairs with low SNR and poor parameter measurements.  
The code behind this pipeline, \gwphase, is publicly available. 

Because we are working directly with the GW phases, we can determine for a given candidate pair the most probable lensing phase shift and hence their image types. This information together with the ability of our technique to rapidly consider all possible pairs makes our method a valuable tool for electromagnetic follow ups, as it could then inform the potential lens model. 
Moreover, due to the good correlation of our distance statistic with the full joint parameter estimation inference (as measured by the coherence ratio, cf. Fig.\ \ref{fig:phase_vs_clu_gwtc}), our method is well positioned to handle the ever-growing number of possible pairs in future catalogs.

In its application for searching for lensed GWs, our method could be extended in several ways. Consistency between a pair could not only be evaluated in terms of the detector phases, but it could also include the polarization state as a function of frequency: this would effectively determine spin consistency. 
In addition, it could be used to forecast the prospects of detecting lensing with future ground- and space-based GW facilities. 
In this respect, our method is expected to improve rapidly with the number of detectors in the GW network, and aspect that we will explore in future work. 

We have also found that the orbital phase evolution is more constraining than the chirp mass, typically by a factor of 2-3 in terms of parameter distance.  For any given threshold in detection or distance this substantially enhances the purity of the catalog and may even be useful for  subthreshold searches. Similarly, the time delay phases offer an interesting opportunity to improve over the largely unlocalized subthreshold triggers. 
In terms of its current implementation, \gwphase\  could be improved with machine learning algorithms to compute the distance in low significance, non-Gaussian posteriors. 

Although we focused on its use to discover lensed GWs, the code developed to post-process GW phases has far reaching applications. 
For example, tracking the phase evolution at different frequencies one could test the theory of gravity, and by looking at the phase difference of the polarization states one could probe birefringence. 
Finally, while we have focused on the dominant quadrupole mode, comparing the phases of different modes can potentially address waveform systematics.

\begin{acknowledgments}
We are grateful to Srashti Goyal for comments on the paper and Geraint Pattern for correspondence about the IMRPhenomX waveform family. 
JME is supported by the European Union’s Horizon 2020 research and innovation program under the Marie Sklodowska-Curie grant agreement No. 847523 INTERACTIONS, and by VILLUM FONDEN (grant no. 53101 and 37766). 
WH is supported by U.S.\ Dept.\ of Energy contract DE-FG02-13ER41958 and the Simons Foundation. 
RKLL is supported by the National Science Foundation through award number PHY-1912594 and PHY-2207758. 
This material is based upon work supported by NSF's LIGO Laboratory which is a major facility fully funded by the National Science Foundation. The authors are grateful for computational resources provided by the LIGO Laboratory and supported by National Science Foundation Grants PHY-0757058 and PHY-0823459.
\end{acknowledgments}

\appendix

\section{Reference frames and polarization basis conventions}
\label{sec:frames}

\begin{figure*}[t!]
\centering
\includegraphics[width = \textwidth,valign=t]{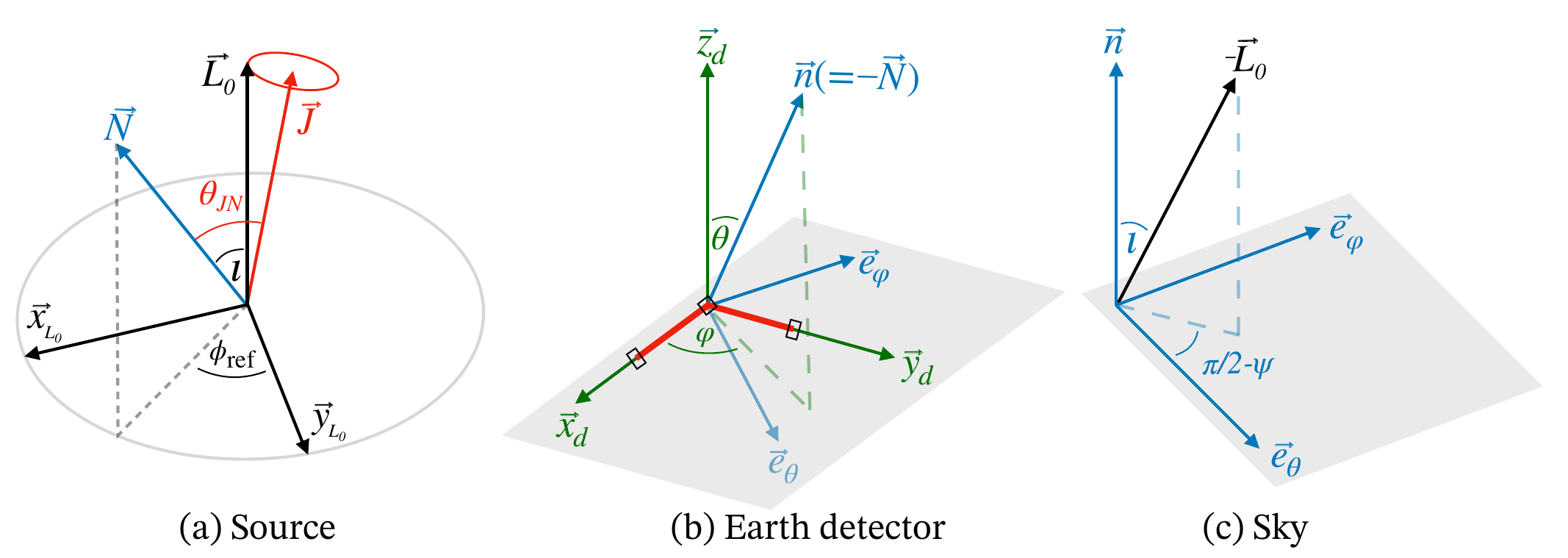}
\caption{Summary figure of our frame conventions. 
The source frame (a) defines the coordinate system in which the intrinsic parameters of the binary are defined: masses, spins and phase. It is anchored to the orbital angular momentum at the reference frequency $\vec L_0$. 
The Earth detector frame (b) serves to define the time of arrival and the position of the sky of the binary event for a fiducial detector at the center of the Earth, in order to compute the antenna response function of each detector. 
The sky frame (c) defines the remaining extrinsic parameters, the inclination $\iota$ and the polarization angle $\psi$. 
See text for further details.}
\label{fig:frame_conventions}
\end{figure*}

The gravitational wave signal emitted during a compact binary coalescence can be defined in different reference frames. To simplify comparisons with existing codes, we will follow the LALSuite conventions for gravitational wave data analysis \cite{lalsuite} (used by LVK, waveform modelers, e.g \cite{Pratten:2020ceb}, and numerical relativists \cite{Schmidt:2017btt}). 

The \emph{source frame} 
defines the reference frame at a given epoch where the intrinsic parameters of the binary are inputted. 
We fix this initial frame to a reference frequency of the GW $f_\rref$    
and anchor it to the orbital angular momentum of the binary at the time the reference frequency is emitted in the 22-mode,  
$\vec{L}_0=\vec{L}(f_\rref)$, which defines the $z$-axis. Therefore, we call the axes of this reference frame $\{\hat x_{L_0}, \hat y_{L_0}, \hat z_{L_0}\}$. 
The $\hat x_{L_0}$-axis connects the compact objects, 
and $\hat y_{L_0} = \hat z_{L_0} \times \hat x_{L_0}$. The propagation direction $\vec N (=-\vec{n})$ pointing from the source to the observer 
has the spherical-polar angles $(\iota(f_\rref),\pi/2-\varphi_\rref)$, related to the inclination and phase at the reference frequency.
The \emph{radiation frame} (also known as wave frame) is defined by $\vec{N}$ with the $x$-axis, which defines the $+$ polarization, along the transverse projection of $\vec{L}_0$.
Note that for a precessing binary the total angular momentum $\vec J$ defines an alternate inertial frame.  

In terms of the observer, the Earth \emph{detector frame} is defined by the coordinate system $\{\hat x_d, \hat y_d, \hat z_d\}$ at a fixed time $t_d$ with the origin at the center of the Earth.\footnote{Note that for current detected sources and detector sensitivity one can assume that the detector frame is fixed throughout the duration of the signal. This will change with next-generation ground-based detectors where the Earth rotation should be taken into account, and certainly for LISA where signals could last years in band.} 
In the detector frame the sky position of the source is $\{\theta, \varphi\}$. 
Conversely, the \emph{sky frame} is defined by $\{\hat e_\theta,\hat e_\varphi,-\vec N\}$ and  
$\pi/2-\psi$ is the angle of the  transverse projection of $\vec L_0$ which describes the orientation of the polarization.

Following these LVK conventions \cite{Anderson:2000yy,ligo_antenna}, the antenna response functions
for a fiducial detector at the center of the Earth with arms in the $\vec{x}_d$ and $\vec{y}_d$ directions are given by 
\begin{align}
    F_+\! &=  -\frac{1+\cos^2\!\theta}{2}\cos(2\varphi)\cos(2\psi) - \cos\theta\sin(2\varphi)\sin(2\psi), \nonumber\\
    F_\times\! &=  \frac{1+\cos^2\!\theta}{2}\cos(2\varphi)\sin(2\psi) - \cos\theta\sin(2\varphi)\cos(2\psi)\,.
\end{align}
It will be convenient for later to introduce the responses to the left and right circular polarization states
\begin{align}\label{eq:def_F}
F_L &= \frac{F_+ +i F_\times}{\sqrt{2}} \equiv |F_L| e^{i\alpha}, \nonumber\\
F_R &= F_L^* = |F_L| e^{-i\alpha},
\end{align}
where $\alpha=\text{arg}(F_L)=\text{atan}[F_+,F_\times]$. 
Notice that under a change in the polarization angle $\psi$ at fixed $(\theta,\varphi)$, $F_L$ and $F_R$ pick up a pure phase of equal and opposite sign.

Since for a precessing binary, spins and inclination are in general frequency dependent, we define them at the source frame fixed by $\fref$ and $\vec L_0$.
Related to the three vectors $\vec J$, $\vec N$ and $\vec L_0$, there are three relevant angles: the angle between $\vec J$ and $\vec N$, $\theta_{JN}$, defined by $\cos\theta_{JN}=\vec{J}\cdot{N}/|\vec{J}||\vec{N}|=N_z$, and the two spherical polar angles of $\vec J$ in the $\vec L_0$ frame $\{\theta_{JL_0},\phi_{JL_0}\}$, that is $\cos\theta_{JL_0}=\vec{J}\cdot{\vec L_0}/|\vec{J}||\vec{L_0}|$ and $\cos\phi_{JL_0}=J_{x_{L_0}}/(J_{x_{L_0}}^2+J_{y_{L_0}}^2)$. 
\footnote{Note that in the GW literature it is common to simply denote the angles $\theta_{JL_0}$ and $\phi_{JL_0}$ as $\theta_{JL}$ and $\phi_{JL}$, specifying that they are defined at the reference frequency.} 
The inclination of the binary $\iota$ is defined by the angle between $\vec L_0$ and $\vec N$. 
The component spins $\vec S_{1}$ and $\vec S_2$ are defined by their 6 Cartesian coordinates. It is common however in parameter estimation to refer instead to their dimensionless spin magnitudes $\chi_1$ and $\chi_2$, their tilt  
with respect to the Newtonian orbital angular momentum $\vec L_N$ (which is always perpendicular to the binary's orbital plane) 
denoted as $\phi_1$ and $\phi_2$, and $\phi_{12}$ the difference in azimuthal angles of $\vec S_1$ and $\vec S_2$.   

To obtain the actual  response of each individual detector we rotate from Earth detector frame  to the individual detector frames given their orientations and the source position in Earth detector coordinates $(\theta,\varphi)$ to $(\text{ra},\text{dec})$ using the Greenwich Mean Sidereal Time of the observation ($\text{ra}=\varphi + \text{gmst}$ and $\text{dec}=\pi/2-\theta$).  
Conveniently, LALSuite has integrated routines to easily change from the Earth detector (also known as Earth fixed) frame to each of the actual detector frames, for our analysis Hanford, Livingston and Virgo.

\section{Measured phase at a new frequency}
\label{sec:phase_new_freq}

The GW signal is constructed following a given waveform model with intrinsic and extrinsic parameters defined at a reference frequency $\fref$. Specifying $\fref$ is necessary as for precessing binaries the spin components vary over time. LVK analyses typically fix $\fref=20$Hz, which for typical signals does not correspond to the frequency of the best measured phase.  
Our goal here is to demonstrate how to consistently reconstruct the detected phase of a GW at a new frequency. This will also serve to illustrate the working procedure of the \gwphase\  package.

\subsection{Polarization states}
\begin{figure}[t!]
\centering
\includegraphics[width = 0.8\columnwidth,valign=t]{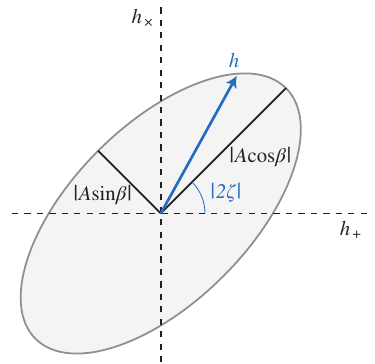}
\caption{A general gravitational wave $h$ is elliptically polarized, where the polarization state rotates between the linear polarization states $h_+$ and $h_\times$ with $\zeta$ defining the orientation of the principal axes and $\beta$ defining their relative amplitude.}
\label{fig:polarization_angles}
\end{figure}

The frequency domain waveform of an emitted GW is described by the real and imaginary parts of the two tensor polarizations.\footnote{As in the main text, we implicitly write all our expressions for positive frequency and left the negative frequencies to be defined by the reality condition of the time domain signal.} 
These polarizations can be described in different bases, see \cite{Isi:2022mbx} for a review, and are fully characterized by four real numbers, two amplitudes and two phases. 
For example, for the linear polarizations of a monochromatic signal this corresponds to $\tilde h_+=A_+e^{-i\phi_+}$ and $\tilde h_\times=A_\times e^{-i\phi_\times}$. 
We follow App.\ C of \cite{Ezquiaga:2021ler} and perform a Stokes decomposition to construct a more physically intuitive set of parameters: $\{A,\, \phi,\, \beta,\, \zeta\}$, where $A$ and $\phi$ describe the global amplitude and phase of the GW signal, whereas the angles $\beta$ and $\zeta$ characterize the relation between the two polarization components. In order to connect the new set of parameters to the ones of the polarizations, it is convenient to move to the circular basis
\begin{align}
    \tilde h_L&=(\tilde h_+-i\tilde h_\times)/\sqrt{2}\,, \\
    \tilde h_R&= (\tilde h_+ + i \tilde h_\times)/\sqrt{2}\,,
\end{align}
which can also be decomposed in an amplitude and a phase $\tilde h_L=A_Le^{i\phi_L}$ and $\tilde h_R=A_R e^{i\phi_R}$\,. 
The global amplitude and phase are defined by
\begin{align} \label{eq:phase}
    \phi=(\phi_L + \phi_R)/2\,,\quad A=\sqrt{A_L^2+A_R^2}\,.
\end{align}
The angle $\beta$ is related to the ratio of the semi-major and minor axes
\begin{equation}
    \tan \beta = \frac{1 -r}{1+r}
\end{equation}
where $r$ is the amplitude ratio $r=A_L/A_R$.  
The difference in the phases describes the orientation of the semi-major axis or equivalently a rotation of the $+,\times$ basis
\begin{equation}
     \zeta = (\phi_L-\phi_R)/4\,.
\end{equation}
The circularly-polarized modes can then be written as
\begin{align}
    \tilde h_L&=h_Rre^{-4i\zeta}=\frac{Ar}{\sqrt{1+r^2}}e^{-i(\phi+2\zeta)}\,,\\
    \tilde h_R&=A_Re^{-i \phi_R}=\frac{A}{\sqrt{1+r^2}}e^{-i(\phi-2\zeta)}\,.
\end{align}
Purely linear polarization corresponds to $\beta=0$ (or $r=1$), with $h_+$ if $\zeta=0$ and $h_\times$ 
if $\zeta=\pm\pi/4$.  Purely circular polarization corresponds to $\beta=\pm \pi/4$. 
A graphical representation of the angles defining the polarization state is shown in Fig.\ \ref{fig:polarization_angles}.

The frequency domain waveform of an emitted GW defined by a given waveform model is outputted in the radiation frame. 
For generic precessing binaries, the orbital angular momentum $\vec L$, polarization angle $\psi$ and inclination $\iota$ all change with time and, therefore, frequency. Similarly, all $\{A,\, \phi,\, \beta,\, \zeta\}$ will be frequency dependent. 
In our default analysis we will use the IMRPhenomXPHM waveform model \cite{Pratten:2020ceb} that was used to analyze the latest GW catalog \cite{LIGOScientific:2021djp}. 

\subsection{Detected phase at a new frequency}

Our main goal is to compute the detected phase of the GW at any frequency, i.e.\ Eq.\ (\ref{eq:phi_d}). 
For this we first need the GW global phase $\phi(f)$ in the radiation frame which can be computed from the left and right phase following Eq.\ (\ref{eq:phase}). 
This phase can be constructed for any multipole mode, provided we decompose the signal in its multiple moments $\tilde h = \sum_{lm}A_{lm}e^{-i\Phi_{lm}}$ to get $\{A_{lm},\, \phi_{lm},\, \beta_{lm},\, \zeta_{lm}\}$.
In our analysis we reconstruct the phase of the 22 mode at different frequencies $\phi_{22}(f)$. 
As noted in the main text, we define the 22-mode in the coprecessing frame from the +-polarization. The equatorial symmetry of this frame together with the reality condition of the time domain signal implies $\tilde h_{lm}=(-1)^l\tilde h^*_{l-m}(-f)$ \cite{Garcia-Quiros:2020qpx}. Therefore only one $m$-mode defines the phase at positive frequencies. In the conventions of the IMRPhenomX waveform family \cite{Garcia-Quiros:2020qpx,Pratten:2020ceb} that we follow, this corresponds to $l=2$, $m=-2$. 

The 22-waveform phase at the reference frequency is then read off directly from the waveform approximant for a given set of binary parameters using Eq.~(\ref{eq:phase}).   We remind the reader that in the stationary phase approximation (SPA)
it is related to the orbital phase  $\phi_\rref$
as
\begin{equation}
    {\left.\Phi_{22}(\fref)\right\vert_\text{SPA}= 2\phi_\rref - 2\pi \fref (t_\rref-t_d) + \mathrm{const}} 
\end{equation}
where $t_\rref$ is the arrival time of the frequency $\fref$. 
We take the global waveform arrival time $t_d$ in Eq.~(\ref{eq:detectedwaveform}) to coincide with this time $t_d=t_\rref$. 
Therefore, our ``detector phase" $\phi_d$ is then the phase of frequency $f$ at the time when $f_\rref$ hits the given detector. 

Given $\phi(f)\equiv \Phi_{22}(f)$,
in order to obtain the detector phase $\phi_d(f)$ we need to project from the radiation to the detector frame. At a given detector, the frequency domain signal of a given $lm$-mode (although we do not include the $lm$ label here for notational simplicity) is
\begin{align}\label{eq:derivation}
\tilde h_d &= F_L \tilde h_L + F_R \tilde h_R 
\nonumber\\
&= 
\frac{A|F_L| }{\sqrt{1+r^2}}e^{-i\phi}\left(re^{+i(\alpha-2\zeta)} + e^{-i(\alpha-2\zeta)}\right) \nonumber\\
&\equiv A_d(f) e^{-i(\phi+\chi)}\,,
\end{align}
where we have used Eq.~(\ref{eq:def_F}), and
\begin{align}
    \tan\chi&= \frac{(1-r)}{(1+r)}\tan(\alpha-2\zeta) 
\end{align}
Finally, noting that the frequency dependent polarization angle $\zeta$ can be reabsorbed into a redefinition of  $\psi'=\psi+\zeta$, we arrive at the detector phase
\begin{align}
    \phi_d(f)&  = \phi(f) + \chi(f) \\
    &=\phi(f) + \text{atan}\left[F_+(\psi'),\frac{(1-r)}{(1+r)}F_\times(\psi')\right]\nonumber
\end{align}
and amplitude
\begin{equation}
    A_d = \frac{A|F_L|}{\sqrt{1+r^2}}\sqrt{1 + r^2 +2r\cos(2\alpha-4\zeta)}\,.
\end{equation}

With this derivation, the role of the circular polarization ratio $r$ and the linear polarization angle $\zeta$ become more transparent.  
After the multipolar decomposition, the ratio of the polarization amplitudes, $r_{lm}$, provides a direct way of computing the inclination at a new frequency
\begin{equation}
    r_{lm}=\frac{|h^{lm}_L|}{|h^{lm}_R|}=\frac{1-a_{lm}(\iota)}{1+a_{lm}(\iota)}\,,
\end{equation}
where $a_{lm}$ gives the relative strength of the linear polarizations in the emission direction. For $l=|m|$ modes one gets
\begin{equation}
    a_{l=|m|}(\iota(f))= \frac{2\cos\iota}{1+\cos^2\iota} = \frac{1-r_{lm}}{1+r_{lm}}\,.
\end{equation}
This shows that a change in the polarization ratio can be mapped to a change in the inclination, and that both quantities are in general frequency dependent. 

\begin{figure}[t!]
\centering
\includegraphics[width = \columnwidth,valign=t]{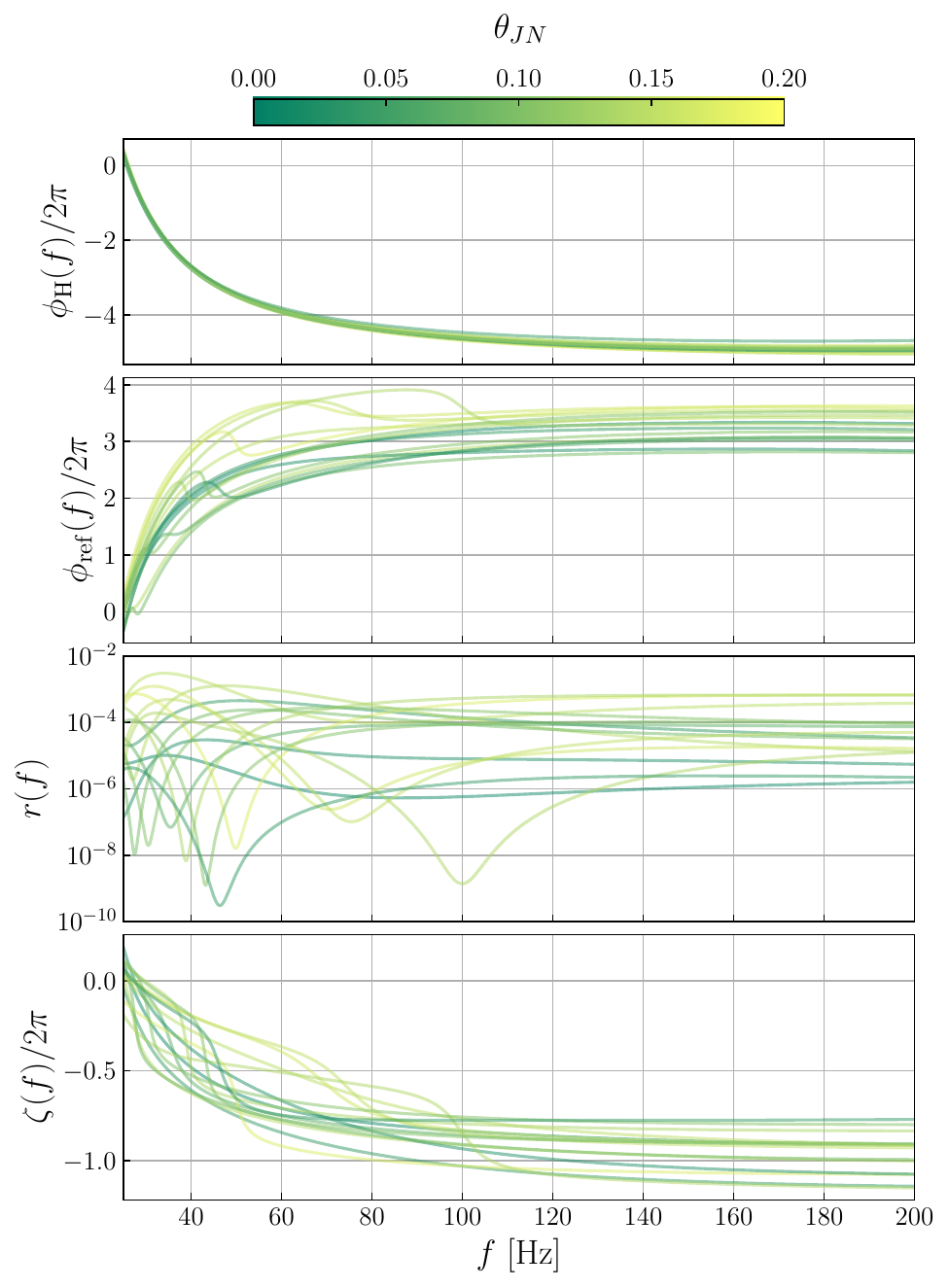}
\caption{
Frequency evolution for the detector phase ($\phiH$) in comparison to the 22 reference phase ($\phi_\text{ref}$), the polarization amplitude ratio ($r$) and the polarization phase difference $\zeta$ for random samples close to face-on. The colors indicate the value of the angle between the total angular momentum $\vec J$ and the line of sight $\vec N$. 
The samples correspond to the simulated Type I, low SNR, HLV event. 
The reference frequency is at 20Hz.
}
\label{fig:polarization_evolution}
\end{figure}

\begin{figure*}[t!]
\centering
\includegraphics[width = \columnwidth,valign=t]{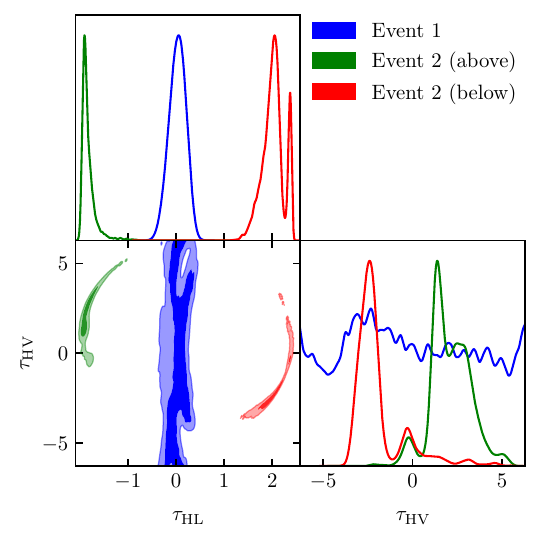}
\includegraphics[width = 0.98\columnwidth,valign=t]{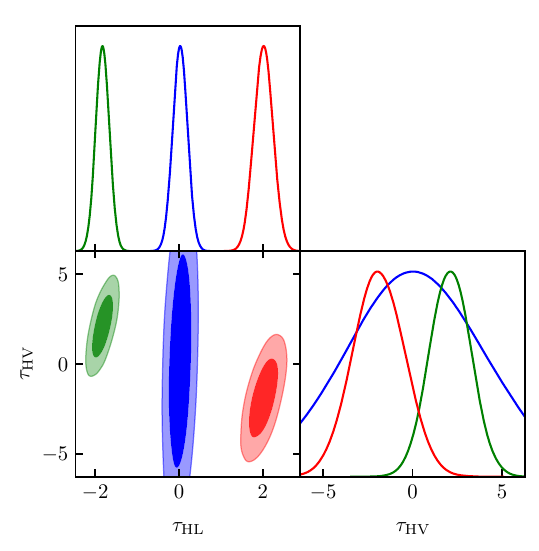}
\caption{ 
Time delay phases for a pair of events analyzed under the lensing hypothesis. Event 1 corresponds to GW190706\_222641 and Event 2 to GW190519\_153544. Event 2 phases are reconstructed at the reference frame determined by Event 1. The above and below samples of Event 2 are defined with respect to its detector plane, i.e.\ $\vec{n}\cdot\left(\vec{r}_{\HL}\times \vec{r}_{\HV}\right)$ negative and positive respectively. 
The right hand plot corresponds to the Gaussianization of the posteriors.}
\label{fig:time_delay_phases}
\end{figure*}

On the other hand, $\zeta$ represents a frequency dependent rotation of the polarization basis as it is clear in Eq.\ (\ref{eq:derivation}). 
Note that by defining the radiation frame $+$ polarization through the projection of $\vec{L}_0$ at the reference frequency $\zeta(f_\text{ref})=0$. 

Finally, let us note that the reconstructed detector phase is typically a well-constrained, monotonic function in frequency even when other quantities such as the inferred reference phase as a function of frequency is not.   This illustrates how our choice of parameters removes degeneracies between binary parameters, in this case the reference phase and inclination.
With precessing binaries, the inclination of the orbital plane can precess from a finite value at the reference frequency to nearly face/edge on at a different frequency and encounter large degeneracies. 
To exemplify this limiting case we draw the frequency evolution for $\phi_\text{ref}$, $r$ and $\zeta$ in Fig.\ \ref{fig:polarization_evolution} for random samples with small $\theta_{JN}\ll1$. 
It is to be noted that the jumps in $\phi_\text{ref}$ and $\zeta$ compensate with $r$ to give a smooth detector phase in the top panel. Here we can also see how the spread in the reference phase is much larger than in the detector phase.

\section{Breaking the bimodality of the time delay phases}
\label{app:breaking_bimodality}

As discussed in the main text,
constant time delay phases track rings on the sky which intersect in points for multiple detector pairs.   When the time delays themselves are well measured, transforming localization to time delays is advantageous as it collapses these degeneracies.  On the other hand, when time delays are derived from other localization information, as is the case for analyzing event 2 in the reference frame of event 1, this collapse does not fully occur and constraints on the derived delays inherit the ring like or multiple point intersections of the localization.  For three detectors this often leads to a bimodal distribution of time delays.
While the information from localization is still retained in the joint time-delays and can be used in parameter distances, bimodality implies non-Gaussianity which degrades our simple Gaussian distance approach. 

This bimodality can however be broken when identifying the parameter samples that come from above or below the plane defined by the three detectors since time delays are symmetric under reflection across this plane. 
Operationally this can be achieved looking at the samples with positive or negative product:
\begin{equation}
    \vec{n}\cdot\left(\vec{r}_{d_1 d_2}\times \vec{r}_{d_1 d_3}\right)\,,
\end{equation}
where $\vec r_{d_i d_j}\equiv \vec r_{d_i} - \vec r_{d_j}$ with $\vec r_{d_i}$ as the position of detector $d_i$.

In Fig.\ \ref{fig:time_delay_phases} we exemplify this construction. We choose GW190706\_222641 as event 1 and GW190519\_153544 as event 2. Event 1 is an HL-detection event, while event 2 is HLV. 
Since three detector events are better localized, it is advantageous to set them as event 2 where the time delay phases encode that information. 
By decomposing the samples of event 2 into those above or below its detector plane we can identify the two independent modes. 
This allow us to asses the (in)consistency of each mode with respect to event 1 by finding the minimum distance between the two. For each of them we can compute the distance using the Gaussian approximation in Eq.\ (\ref{eq:gaussian_distance}). 
If we were not to break the reflection symmetry, the Gaussian approximation would determine that these two events are consistent since its covariance would be forced to cover both modes.

For reference, in the right hand side of Fig.\ \ref{fig:time_delay_phases} we present the Gaussianization of the same time delay posteriors that is used to compute the distance $D_J$. This also serve to demonstrate how the Gaussianization procedure tends to be conservative with respect to rejecting not-lensed pairs.  Here the thin arcs in the joint time delay phases are converted to full ellipses which then are consistent with a wider range of joint time delays.

Since event 1 and event 2 both have this hemisphere/planar degeneracy, when in our method described in \S \ref{sec:Method} we consider both orderings of the events, we are then taking into account the four sections of the sky that the two orientations of the detector planes define.  If in either ordering both modes of the pair is inconsistent with lensing then the lensing hypothesis is rejected.

\section{SNR scaling of phase errors}
\label{sec:leading_order}

In order to estimate the errors in the detected phase and its correlation with other parameters one can use the Fisher information matrix, which is a good approximation in the limit of high signal-to-noise (SNR). If we define the noise-weighted inner product
\begin{equation}
    (a|b)=4\mathrm{Re}\left[\int\frac{a\cdot b^*}{S_n(f)}\text{d}f\right]\,,
\end{equation}
with $S_n(f)$ being the one-sided noise power spectrum density of the noise, we can compute the Fisher matrix
\begin{equation}
    F_{ij}=\left(\frac{\partial h}{\partial \theta_i}\middle\vert\frac{\partial h}{\partial \theta_j}\right)\,,
\end{equation}
where $\theta$ is the set of intrinsic $\theta_\mathrm{int}$ and extrinsic $\theta_\mathrm{ext}$ parameters describing the signal $h$ and the derivatives are evaluated at the true parameters $h=h_T(\theta)$. The covariance matrix is then simply $F^{-1}$. The diagonal of this matrix give us a measure of the standard deviation $\sigma_{\theta_i}=\sqrt{(F^{-1})_{ii}}$. 
For a given model $h_T$ given detector $d$, 
\begin{equation} \label{eq:optimal_snr}
    \rho_d = \sqrt{(h_T|h_T)}\,,
\end{equation}
known as the ``optimal SNR".

To estimate the errors in phase-like and amplitude-like parameters, we use a simple two-parameter toy model that can be thought to describe the signal at a fixed frequency
\begin{equation}
    h_T=\mathcal{A}e^{i\phi}\,.
\end{equation}
From here it is easy to derive that the phase and amplitude will be uncorrelated with standard deviation
\begin{equation}
    \sigma_{\phi}=\sigma_{\mathcal{A}}/\mathcal{A}=1/\rho_d\,.
\end{equation}

In general, the signal will be chirping, and one needs to vary over the reference time and the intrinsic parameters describing the phase evolution like chirp mass. At leading post-Newtonian order for the quadrupolar radiation of a quasi-circular inspiralling binary this can be done analytically \cite{Finn:1992xs,Cutler:1994ys}. Then one finds that the phase is correlated with the time of arrival and the chirp mass, enlarging the errors by a factor of a few. As discussed in the main text, GW150914 has, for example, $\sigma_{\phiH}(40\text{Hz})\simeq 3/\rho_\mathrm{H}$. The precise number depends on the frequency domain signal and the detector's noise. Importantly, heavier binaries will have more weight on their merger and ring-down, making the measured parameter errors larger for a given $\rho_d$ and this simple inspiral approximation less reliable. Similarly, signals that do not fit well the data where $(h|h)$ differs substantially from $(h_T|h_T)$, for example if the precession is not properly modelled or glitches remain in the data, can downgrade the errors of the 22-phase. 

Moreover,  Fisher estimates only include statistical errors whereas our parameter based phase inferences include all errors that are modelled in the original binary parameter inference. LVK analyses marginalize over calibration uncertainties, enlarging amplitude and phase errors. Calibration upper limit errors in LIGO detectors during the third observing run are $<0.07$ radians ($<4$ degrees) within 20-2000 Hz at 68\% CL, although generically they are $\sim0.02$ radians \cite{Sun:2020wke,Sun:2021qcg}. In the first and second observing run LIGO detector had an upper calibration error $\lesssim 0.05$ radians ($\lesssim 3$ degrees) \cite{LIGOScientific:2017aaj}. Note however that these errors are correlated across different frequencies and our phase inference at any given frequency combines information from all frequencies.

\begin{figure*}[t!]
\centering
\includegraphics[width = \textwidth,valign=t]{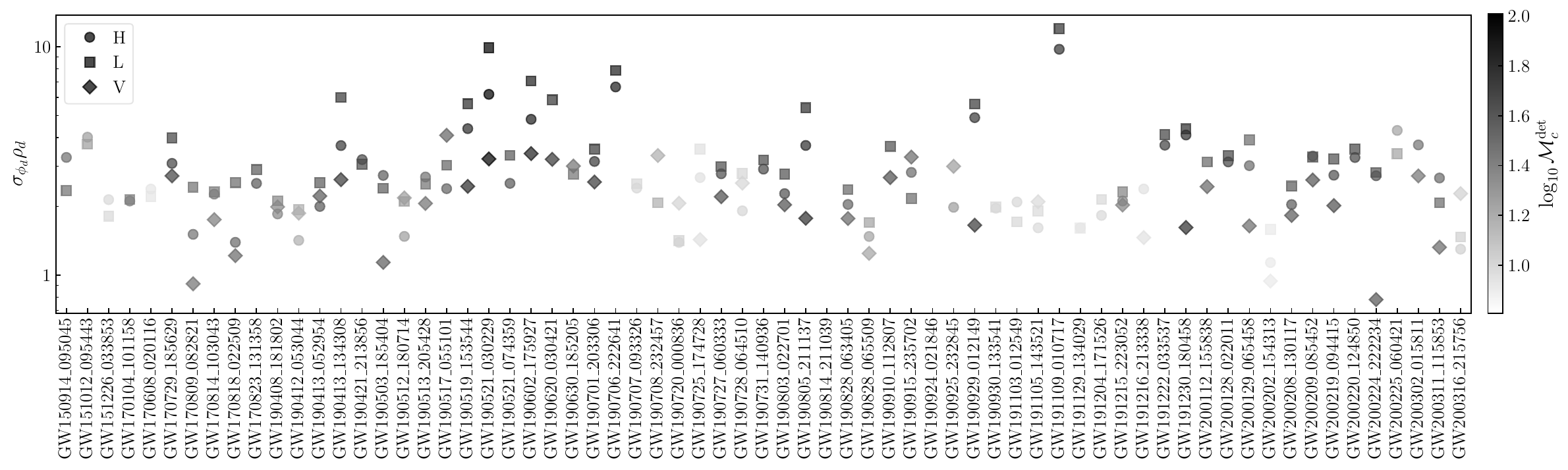}
\caption{Detector phase standard deviation $(\sigma_{\rho_d})$ times median ``optimal" detector SNR ($\rho_d$) for the binary black holes analyzed in this work (GWTC3 with $p_\text{astro}>0.8$ and $\rho_\text{ntw}>8$). The color bar indicates the median detector-frame chirp mass. Detector phases are at 40Hz.}
\label{fig:detector_error_phases_gwtc}
\end{figure*}

For reference, we present in Fig.\ \ref{fig:detector_error_phases_gwtc} the product of the standard deviation of the detector phase and the (median) optimal SNR over the parameter posteriors of the events considered in our analysis. 
The full posterior distribution of the detector phases is given in Fig.\ \ref{fig:detector_phases_gwtc}. 
One can see how for most of the events the error is $\sim\text{few}/\rho_d$, although there are some outliers with larger errors. 
It is to be noted that, GW191109\_010717, the heaviest event in the second half of the third observing run has the largest error. However this event had glitches overlapping with the signal in both detectors and has been shown to keep some anomalies even after removing those, see App.\ A.3.\ in \cite{LIGOScientific:2021sio}.

\begin{table*}
\centering
\begin{tabular}{llccccc}
\hline
\hline
Injection & Characteristics & Network & $\rho_\mathrm{H}$ & $\rho_\mathrm{L}$ & $\rho_\mathrm{V}$ & $\rho_\text{ntw}$ \\
\hline
\hline
1 & Type I, low SNR & HL & $9.0$ & $12.3$ & - & $15.2$\\
2 & Type II, low SNR & HL & $6.7$ & $9.1$ & - & $11.3$\\
3 & Not-Lensed, low SNR & HL & $8.0$ & $8.9$ & - & $12$\\
4 & Type I, low SNR & HLV & $9.0$ & $12.3$ & $6.9$ & $16.7$\\
5 & Type II, low SNR & HLV & $6.7$ & $9.1$ & $4.8$ & $12.3$\\
6 & Not-lensed, low SNR & HLV & $8.0$ & $8.9$ & $3.3$ & $12.4$\\
7 & Type I, high SNR & HL & $17.2$ & $23.5$ & - & $29.1$\\
8 & Type II, high SNR & HL & $12.8$ & $17.4$ & - & $21.6$\\
9 & Not-lensed, high SNR & HL & $15.2$ & $16.9$ & - & $22.7$\\
10 & Type I, high SNR & HLV & $17.2$ & $23.5$ & $13.1$ & $31.9$\\
11 & Type II, high SNR & HLV & $12.8$ & $17.4$ & $9.1$ & $23.4$\\
12 & Not-lensed, high SNR & HLV & $15.2$ & $16.9$ & $6.2$ & $23.6$\\
13 & Type I, low SNR, low mass & HLV & $9.5$ & $11.1$ & $9.9$ & $17.6$\\
14 & Type II, low SNR, low mass & HLV & $7.1$ & $9.6$ & $6.1$ & $13.4$\\
15 & Not-lensed, low SNR, low mass & HLV & $7.5$ & $10.0$ & $5.0$ & $13.5$\\
16 & Type I, high SNR, new $\theta_{JN}$ & HLV & $16.8$ & $23.8$ & $13.4$ & $32.1$\\
17 & Type II, high SNR, new $\theta_{JN}$ & HLV & $12.7$ & $17.8$ & $9.3$ & $23.8$ \\
\hline
\end{tabular}
\caption{\label{table:injections_snrs}
Summary of the true optimal signal-to-noise ratio ($\rho_d$) in each detector and the network SNR for all simulated events. Each injection has a different noise realization. 
The network SNR is computed as $\rho_{\rm ntw}^2= \rho_\mathrm{H}^2 + \rho_\mathrm{L}^2 + \rho_\mathrm{V}^2$.}
\end{table*}

\section{Details on the simulated gravitational waves}
\label{app:injections}

In this appendix we detail all the configuration settings for the simulated events. 
A table with the parameters of the injections is presented in Table \ref{table:injections_parameters}, where we follow the standard parameter definition of \texttt{bilby} \cite{Ashton:2018jfp}. 
For triggers 1, 2 and 3 we perform injections for both HL and HLV detector configurations. 
For the rest of injections we only set HLV configurations. 
In total we analyze 17 injections. 
In all our analyses we use the projected sensitivites for the fourth observing run as described by the amplitude spectral densities \texttt{aligo\_O4high} for both LIGO detector and \texttt{avirgo\_O4high\_NEW} for Virgo in the public LIGO Document T2000012-v1 (\href{https://dcc.ligo.org/LIGO-T2000012-v1/public)}{https://dcc.ligo.org/LIGO-T2000012-v1/public}). 
The optimal  SNR (for the true injected values) for all the injections is summarized in Table \ref{table:injections_snrs}. 

\begin{figure*}[t!]
\centering
\includegraphics[width = \textwidth,valign=t]{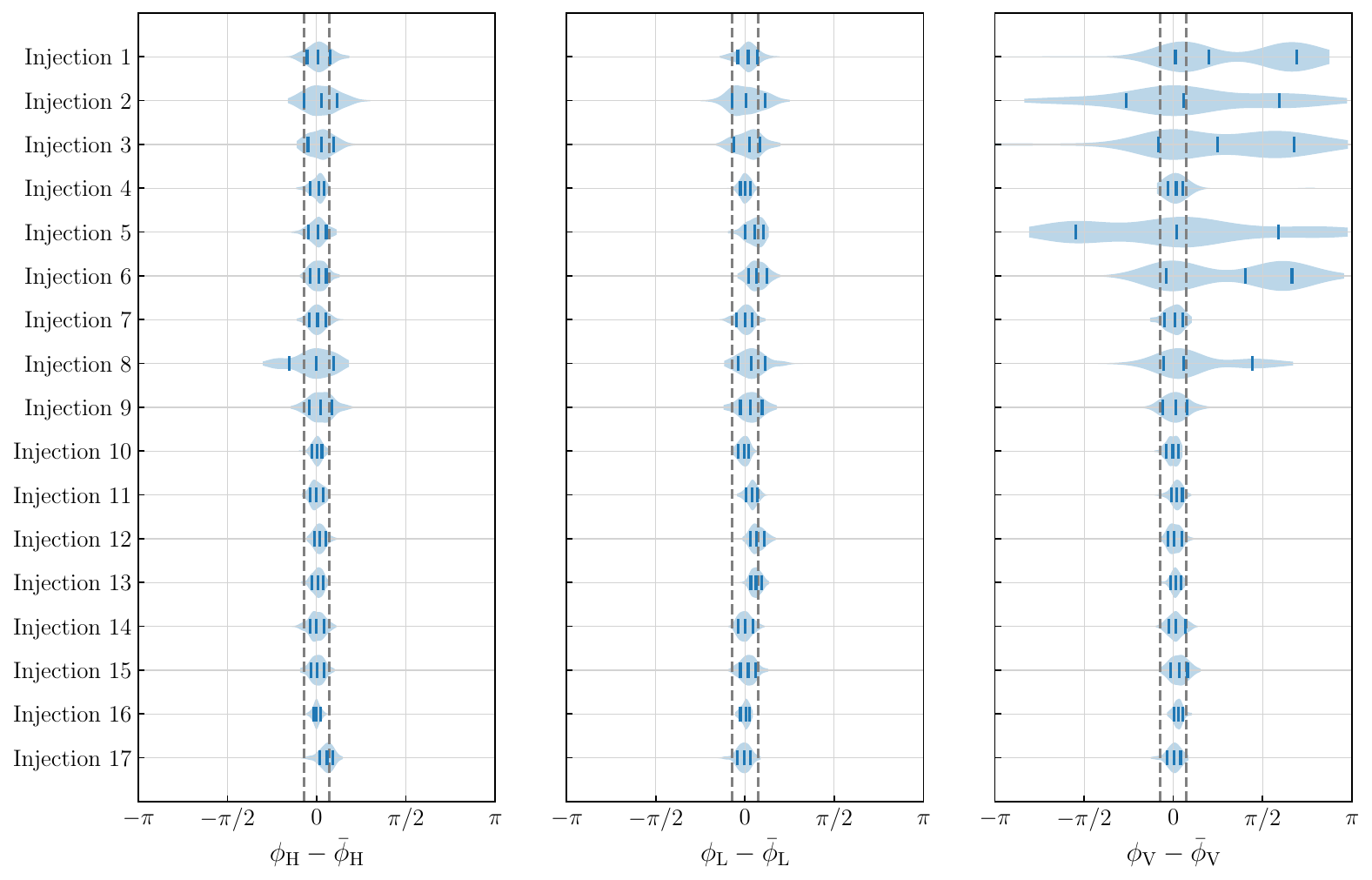}
\caption{Posterior distribution of the detector phases centered at their median value for the set of simulated GW events. The blue ticks indicate the $68\%$ credible interval and the dashed vertical lines indicate the prior width for an informative phase of $\pi/2\sqrt{12}$, see main text for details.}
\label{fig:detector_phases_injections}
\end{figure*}

In Fig.\ \ref{fig:detector_phases_injections} we present the posterior distributions for the detector phases of all the injections. 
For most of the cases, except the HL detections, the detector phases $68\%$ CL is comparable or smaller than $\pi/2\sqrt{12}$, indicating that they are informative against the lensing hypothesis.

\begin{table*}
\centering
\begin{tabular}{lcccccccccccccccc}
\hline
\hline
Injection & $m_1\,[M_\odot]$ & $m_2\,[M_\odot]$ & $a_1$ & $a_2$ & $\phi_1$ & $\phi_2$ & $\phi_{12}$ & $\phi_{JL}$ & $\theta_{JN}$ & ra & dec & $\psi$ & $\phi_\text{ref}$ & $d_L\,$[Mpc] & $t_\text{ref}\,$[sec] & Morse phase \\
\hline
\hline
1, 4 & 35.6 & 30.6 & 0.2 & 0.1 & 0.6 & 0.3 & 1.2 & 0.5 & 0.8 & 1.0 & 0.52 & 0.7 & 2.0 & 1900.0 & 0.0 & 0\\
2, 5 & 35.6 & 30.6 & 0.2 & 0.1 & 0.6 & 0.3 & 1.2 & 0.5 & 0.8 & 1.0 & 0.52 & 0.7 & 2.0 & 2650.0 & 517988 & $\pi/2$ \\
3, 6 & 35.6 & 30.6 & 0.2 & 0.1 & 0.6 & 0.3 & 1.2 & 0.5 & 0.8 & 0.3 & 0.18 & 0.5 & 3.77 & 2900.0 & 3451153 & 0 \\
7, 10 & 35.6 & 30.6 & 0.2 & 0.1 & 0.6 & 0.3 & 1.2 & 0.5 & 0.8 & 1.0 & 0.52 & 0.7 & 2.0 & 1000.0 & 0.0 & 0\\
8, 11 & 35.6 & 30.6 & 0.2 & 0.1 & 0.6 & 0.3 & 1.2 & 0.5 & 0.8 & 1.0 & 0.52 & 0.7 & 2.0 & 1394.0 & 517988 & $\pi/2$ \\
9, 12 & 35.6 & 30.6 & 0.2 & 0.1 & 0.6 & 0.3 & 1.2 & 0.5 & 0.8 & 0.3 & 0.18 & 0.5 & 3.77 & 1526.0 & 3451153 & 0 \\
13 & 15.4 & 12.6 & 0.2 & 0.1 & 0.6 & 0.3 & 1.2 & 0.5 & 0.8 & 1.0 & 0.52 & 0.7 & 2.0 & 823.0 & 79121.0 & 0 \\
14 & 15.4 & 12.6 & 0.2 & 0.1 & 0.6 & 0.3 & 1.2 & 0.5 & 0.8 & 1.0 & 0.52 & 0.7 & 2.0 & 1147.87 & 428696 & $\pi/2$ \\
15 & 15.4 & 12.6 & 0.2 & 0.1 & 0.6 & 0.3 & 1.2 & 0.5 & 0.8 & 1.41 & 0.54 & 0.5 & 3.77 & 1256.16 & 4918029 & 0 \\
16 & 35.6 & 30.6 & 0.2 & 0.1 & 0.6 & 0.3 & 1.2 & 0.5 & 2.34 & 1.0 & 0.52 & 0.7 & 2.0 & 1000.0 & 0.0 & 0\\
17 & 35.6 & 30.6 & 0.2 & 0.1 & 0.6 & 0.3 & 1.2 & 0.5 & 2.34 & 1.0 & 0.52 & 0.7 & 2.0 & 1394.0 & 517988 & $\pi/2$ \\
\hline
\end{tabular}
\caption{\label{table:injections_parameters}
Summary of the physical parameters for the injections described in Table \ref{table:injections_snrs}: detector frame component masses ($m_{1,2}$), dimensionless spin magnitudes ($a_{1,2}$), the azimuthal angle of the spin vectors ($\phi_{1,2}$), the difference between the azimuthal angles of the individual spin vectors ($\phi_{12}$), the difference between total and orbital angular momentum azimuthal angles ($\phi_{JL}$), the angle between the total angular momentum and the line of sight ($\theta_{JN}$), right ascension (ra), declination (dec), polarization angle ($\psi$), phase at reference frequency ($\phi_\text{ref}$), luminosity distance ($d_L$) and reference time ($t_\text{ref}$). The reference frequency is at $20$Hz. Injections with Morse phase equal to 0, $\pi/2$ and $\pi$ correspond to type I, II and III images respectively.}
\end{table*}

\section{Details on the gravitational waves analyzed}
\label{app:gwtc}

\begin{figure*}[t!]
\centering
\includegraphics[width = 0.9\textwidth,valign=t]{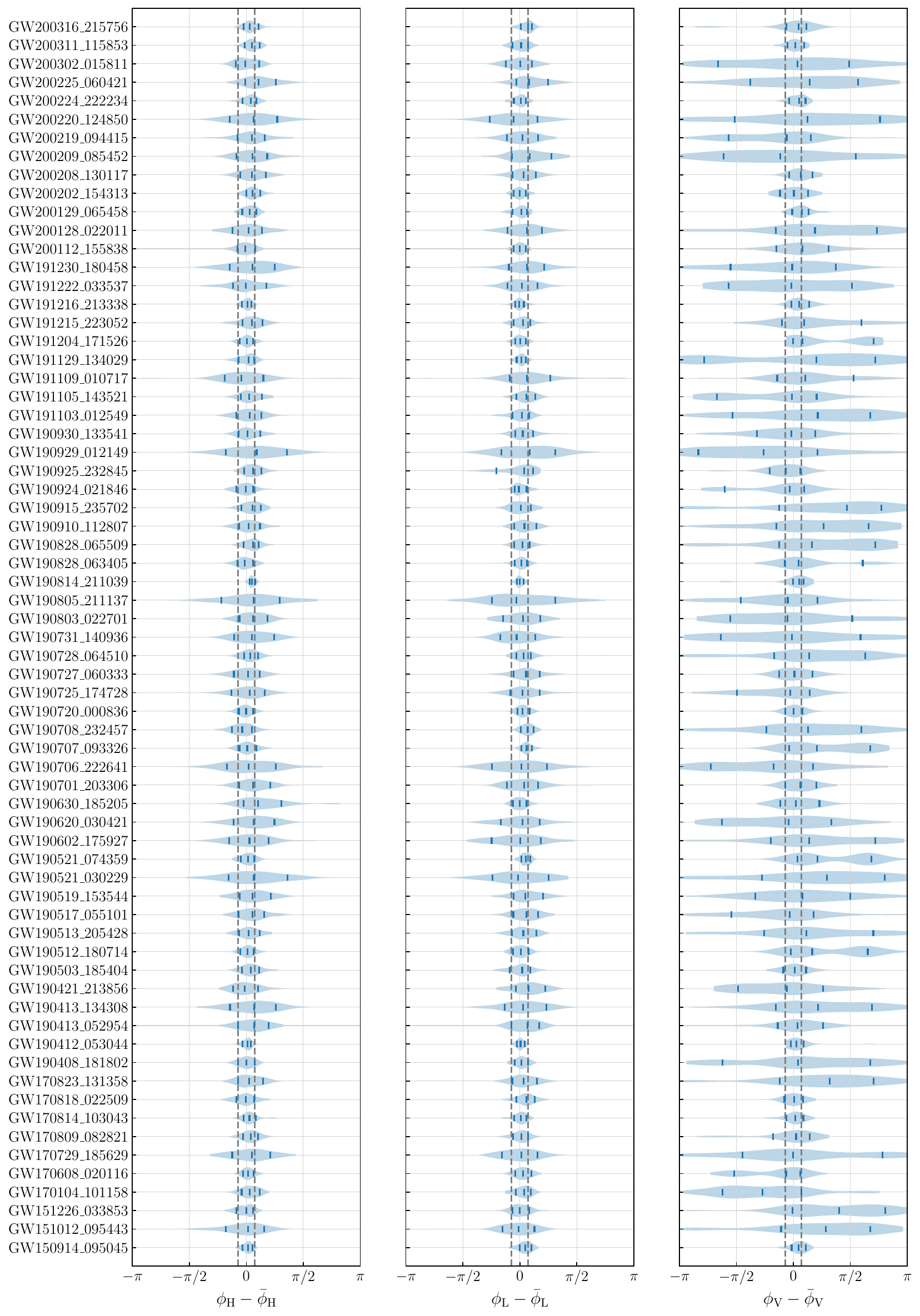}
\caption{Posterior distribution of the detector phases centered at their median value for the GW events analyzed in this paper: $p_\text{astro}>0.8$ and $\rho_\text{ntw}>8$ in GWTC3 \cite{LIGOScientific:2021djp}. 
The blue ticks indicate the $68\%$ credible interval and the dashed vertical lines indicate the prior width for an informative phase of $\pi/2\sqrt{12}$ as in Fig.\ \ref{fig:detector_phases_injections}.}
\label{fig:detector_phases_gwtc}
\end{figure*}

To complement the analyses of the main text, in Fig.\ \ref{fig:detector_phases_gwtc} we present the posterior distribution of the detector phases for all the real events that we have analyzed: 67 binary black holes with $p_\text{astro}>0.8$ and $\rho_\text{ntw}>8$ in GWTC3 \cite{LIGOScientific:2021djp}. 
Following the LVK convention, we use the full name for all the events: UTC date with the time of the event given after the underscore. 

From Fig.\ \ref{fig:detector_phases_gwtc} one can realize that the Hanford and Livingston phases are typically within a fraction of a radian, while Virgo phases are less constrained and in many cases cover the whole prior range. 

For the calculation of the posterior overlap, we follow \cite{Haris:2018vmn,Hannuksela:2019kle,LIGOScientific:2021izm,LIGOScientific:2023bwz} and compute the overlap in $\{m_{1z},m_{2z},\text{ra},\sin\text{dec},a_1,a_2,\cos\phi_1,\cos\phi_2,\cos\theta_{JN}\}$ making the following choices for the prior ranges: $m_{1z,2z}\subset[2,200]M_\odot$, $\text{ra}\subset[0,2\pi]$, $\sin\text{dec}\subset[-1,1]$, $a_{1,2}\subset[0,1]$, $\cos\phi_{1,2}\subset[-1,1]$ and $\cos\theta_{JN}\subset[-1,1]$.  

\bibliographystyle{apsrev4-1}
\bibliography{gw_lensing}
\end{document}